\documentclass{jfm}

\usepackage{graphicx}
\usepackage{bm}
\usepackage{pifont}
\usepackage{mathtools}
\usepackage{newtxtext}
\usepackage{newtxmath}
\usepackage{natbib}
\usepackage{hyperref}
\hypersetup{
    colorlinks = true,
    urlcolor   = blue,
    citecolor  = black,
}

\newcommand{\RomanNumeralCaps}[1]
\linenumbers

\newcommand{\im}{\mathrm{i}}
\newcommand{\e}{\mathrm{e}}
\newcommand*\de{\mathop{}\!\mathrm{d}}
\newcommand{\Oh}{O}
\renewcommand*{\Re}{\operatorname{Re}}
\renewcommand*{\Im}{\operatorname{Im}}
\newcommand*{\ie}{\emph{i.e.}}
\renewcommand*{\eg}{\emph{e.g.}}
\newcommand{\w}{w}
\newcommand{\hatE}{\hat{\mathcal{E}}}
\newcommand{\hattau}{\hat{t}}
\newcommand{\hattaub}{\bm{\hat{t}}}
\newcommand{\hatF}{\hat{F}}
\newcommand{\hatT}{\hat{T}}
\newcommand{\n}{\hat{\nu}}
\newcommand{\s}{\hat{s}}
\newcommand{\F}{\mathcal{F}}

\newcommand{\myeqref}[2]{(\hyperref[#1]{\ref*{#1}#2})}
\newcommand{\myref}[2]{\hyperref[#1]{\ref*{#1}(#2)}}
\newcommand{\myrefnb}[2]{\hyperref[#1]{\ref*{#1}#2}}
\newcommand{\subtag}[1]{\tag{\theequation #1}}

\title{A nematic liquid crystal with an immersed body: equilibrium, stress, and paradox}

\author{Thomas G.~J.~Chandler\aff{1}
  \corresp{\email{tgchandler@wisc.edu}},
Saverio E.~Spagnolie\aff{1}}

\affiliation{\aff{1}Department of Mathematics, University of Wisconsin--Madison, Madison, WI 53706.}

\begin{document}
\maketitle

\begin{abstract}
We examine the equilibrium configurations of a nematic liquid crystal with an immersed body in two-dimensions. A complex variables formulation provides a means for finding  analytical solutions in the case of strong anchoring. Local tractions, forces, and torques on the body are discussed in a general setting. For weak (finite) anchoring strengths, an effective boundary technique is proposed which is used to determine asymptotic solutions. The energy-minimizing locations of topological defects on the body surface are also discussed. A number of examples are provided, including circular and triangular bodies, and a Janus particle with hybrid anchoring conditions. Analogues to classical results in fluid dynamics are identified, including d'Alembert's paradox, Stokes' paradox, and the Kutta condition for circulation selection.
\end{abstract}

\begin{keywords}
Authors should not enter keywords on the manuscript, as these must be chosen by the author during the online submission process and will then be added during the typesetting process (see \href{https://www.cambridge.org/core/journals/journal-of-fluid-mechanics/information/list-of-keywords}{Keyword PDF} for the full list).  Other classifications will be added at the same time.
\end{keywords}


\section{Introduction}\label{sec:intro}

A liquid crystal (LC) is a state of matter in which molecules or elongated composites immersed in a solvent possess orientational order but not positional order \citep{degennes1993}. The local molecular orientation is represented as a director field $\bm{n}(\bm{x})$, with spatial position $\bm{x}$ and $|\bm{n}|=1$. Among the unusual features of a fluid so composed is the elastic response to deformation of the orientation field, which results in a non-vanishing stress on confining or internal boundaries even at equilibrium. The nature of this stress depends critically upon the preferred director orientation on a given surface and the associated anchoring energy. Common boundary conditions include homogeneous (tangential) and homeotropic (perpendicular) alignment \citep{Jerome91}.

Although difficult to obtain in general, an analytical representation of the equilibrium state of a liquid crystal with an internal boundary is greatly desired for theoretical efforts to understand a wide range of phenomena, including LC-mediated elastic interactions between bodies \citep{pslw97,ca08,lms09,trahd12} the deformation of soft immersed bodies \citep{md13,mpwsa16,zzmhap16,nesa20,sv22}, and the dynamics of bodies immersed in active suspensions \citep{rzd23}. Such representations may also be needed to explore the anisotropic viscous drag on moving bodies \citep{rt95,sv01,lhp04,gd13} and in applications like microrheology \citep{gd16,csmsd16}. More detailed discussions have recently been provided by \cite{Muvsevivc17} and \cite{Smalyukh18}. Biological fluids like mucus and dense cell populations also exhibit anisotropy \citep{vhv93,vcht08}. The stresses and dynamics of bodies immersed in these settings may have profound implications for biological function. Among other consequences, fluid anisotropy can strongly impact bacterial locomotion \citep{mttwa14,zsla14,tmasw15,ksp15,fdoa19,su23}. Active colloids and droplets in liquid crystals have accordingly become an appealing synthetic means of examining motile living systems \citep{jgwtlcy15,bbmwa16,Lavrentovich16,kkbm16,nca19}. 

 Adding to the challenge of determining equilibrium states are singularities in the director field, which must appear as a consequence of topological conservation laws \citep{acmk12}; for example, bipolar surface `boojum' singularities can appear in the case of strong homogeneous anchoring \citep{vl83}. Homeotropic anchoring, meanwhile, encourages the appearance of a defect in the bulk LC as either a point singularity, a Saturn-ring line defect \citep{Terentjev95,lpcs98,ga00}, or an even more exotic configuration \citep{lmsc10,fltm14}. The large energy associated with topological defects results in effective interactions between them \citep{lpcs98,gsv02,hs20}, which has been a major recent focus in the dynamics of active suspensions \citep{gbmm13,klsdgbmdb14,diys18,Aranson19,ssbmv22}. In general, the location of the defects on the surface or in the fluid depends on the relationship between the bulk elastic energy and the surface anchoring strength, and while also not simple to determine, is an important part of understanding the LC equilibrium and any associated surface stresses.

Assuming small gradients in the director field, the distortion free energy density within the bulk of a nematic LC is given by
\begin{equation}\label{eq:full_freeenergy}
   \mathcal{F}_\mathrm{bulk} \coloneqq \frac{1}{2} K_1 (\nabla\cdot \bm{n})^2  +\frac{1}{2} K_2 (\bm{n}\cdot\nabla\times\bm{n})^2 +\frac{1}{2} K_3 |\bm{n}\times \nabla\times\bm{n}|^2,
\end{equation}
where  $K_1$, $K_2$, and $K_3$ are the splay, twist, and bend Frank elastic constants, respectively \citep{degennes1993,stewart2004}. A nematic state has only orientational order --- phases with additional structure, for instance smectics with layers of oriented regions, have different energy densities. Here already, though, the free energy in \eqref{eq:full_freeenergy} is frequently too complex to be of practical use since the relative values of the elastic constants, $K_i$, may be unknown, and the corresponding equilibrium equations are generally difficult to solve \citep{degennes1993}. It is thus common to assume the one-constant approximation, \ie~$K\coloneqq K_1=K_2=K_3$, as we shall for the remainder of the paper.

Settings in which the nematic distortions are confined to a plane are particularly amenable to mathematical analysis. The two-dimensional director field can then be described by a director angle, $\theta(x,y)$, which is defined so that $\bm{n}\equiv(\cos\theta,\sin\theta,0)$. (Note that $\theta(x,y)$ is only uniquely defined modulo $\pi$ due to the assumed reflective symmetry of the LC constituents.) With this additional simplification, the free-energy is a Dirchlet energy,
\begin{equation}\label{eq:1const_energy}
 \mathcal{F}_\mathrm{bulk} = \frac{1}{2}K |\nabla\theta|^2.
\end{equation}
Applying the principle of virtual work yields Laplace's equation, $\nabla^2\theta=0$,
which describes the bulk equilibrium configuration of the two-dimensional nematic assuming the one-constant approximation. Thus, were it not for nonlinear mixed boundary conditions and additional constraints imposed by topology, the equilibrium state would be simple to deduce.

While analytical solutions are sparse, some geometries admit analytical forms for these LC configurations for instance,  in or outside tactoid geometries \citep{pv21}, outside a cylinder \citep{burylov1994,ls22}, and in an annulus \citep{lahm17}. Among the more elegant approaches to solving such problems, however, is the use of complex variables and related techniques. This approach has a very long history in fluid mechanics and in a multitude of engineering disciplines \citep{Acheson1990,af03,Crowdy20}. Some recent applications in fluids have included microorganism swimming \citep{clslh11}, dissolution and melting \citep{rb16}, and flow over a bubble mattress \citep{Schnitzer16,yc20}. In the context of liquid crystals, this approach has already been used successfully to study LC configurations in rectangular wells and steps \citep{dm12}, near boundaries with periodic wavy surfaces \citep{dm12,ltlr16}, in two-dimensional spindle shapes \citep{vov12,lmbm17}, and on curved surfaces \citep{vn06}.

In this paper, we set out to determine the equilibrium configuration of a nematic LC outside an immersed body in two-dimensions. To achieve this, we shall utilize a complex variables formulation, which will allow the construction of analytical representations of the director field using techniques such as conformal mapping. We shall also develop an `effective boundary technique', which will enable the consideration of both strong (infinite-strength) and weak (finite-strength) anchoring of the nematic director field to  the immersed boundary alike. Defect locations, net free energies, local surface tractions, and body forces and torques shall all be discussed in the context of this formulation. Furthermore, by considering different body shapes and anchoring conditions, we make analogies with potential fluid flows and related paradoxes. Namely, we will arrive at a paradox similar to the classical d'Alembert paradox, regarding the net force and torque on an immersed body; one similar to Stokes' paradox, regarding the unsolvability of the two-dimensional problem when a non-zero body torque is imposed; and a result similar to the Kutta condition, which selects a free circulation when a body has a sharp corner.

This paper is organized as follows. The mathematical formulation is presented in \S\ref{sec:formulation}, including a discussion of the anchoring boundary conditions and surface traction. The system is then made dimensionless, a complex variables formulation is introduced, and the total force and torque on the body are written as contour integrals. In \S\ref{sec:moderatetangent}, an effective boundary technique is constructed in order to solve problems with finite anchoring strength, and to show that finite anchoring strengths regularize an otherwise unbounded energy associated with topological defects. Then, beginning with \S\ref{sec:ex1}, we present three worked examples, which demonstrate the methodology for determining the two-dimensional director field. In \S\ref{sec:ex1} and \S\ref{sec:ex2}, we consider two domains with homogeneous anchoring conditions: an immersed cylinder and an immersed equilateral triangle. Finally, in \S\ref{sec:ex3}, we use the methodology to solve the problem of a mobile Janus particle with hybrid anchoring conditions. We  close with final remarks in \S\ref{sec:conc}.

\section{Mathematical formulation}\label{sec:formulation}

\begin{figure}
    \centering
    \includegraphics[width=0.8\textwidth]{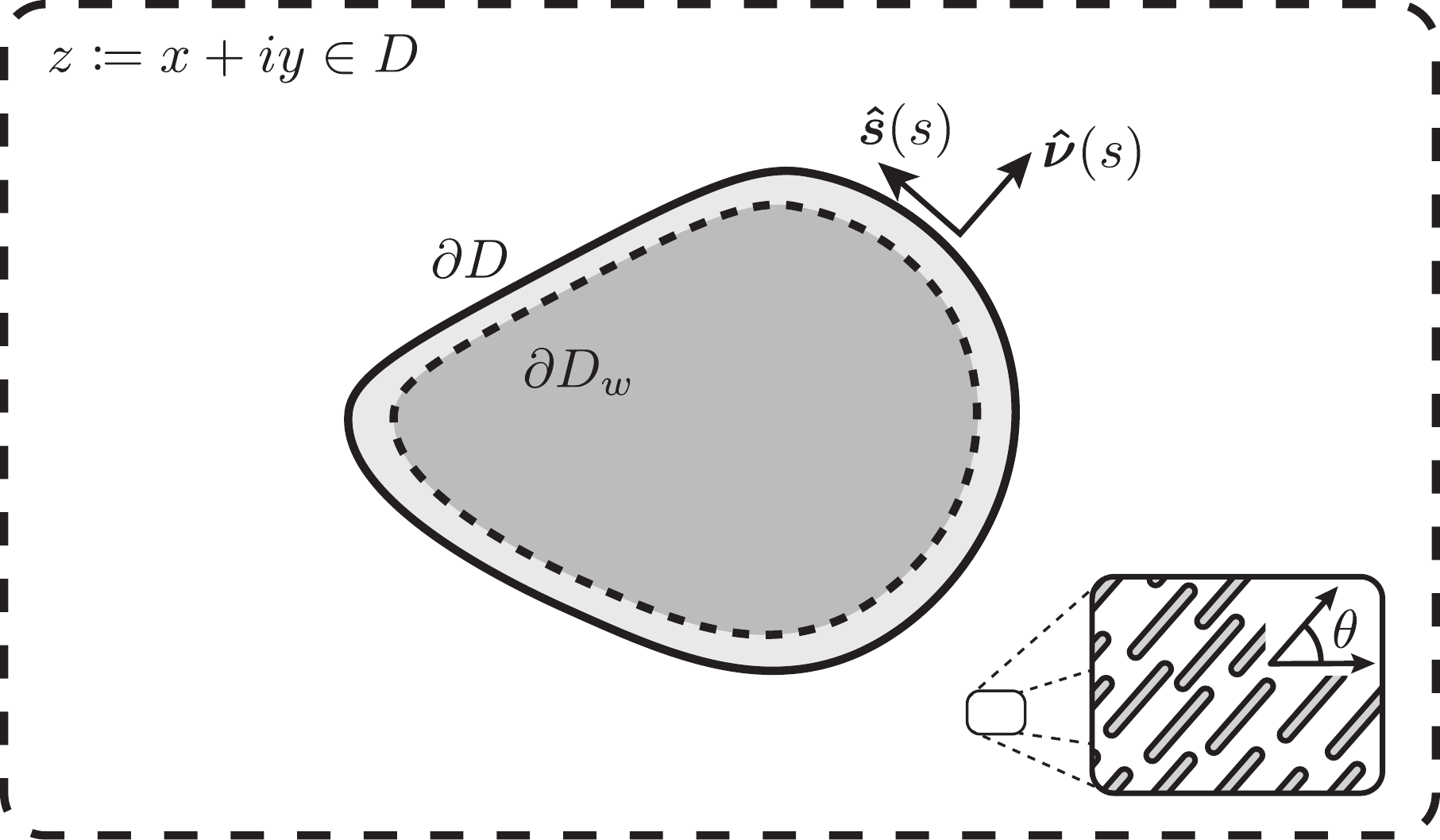}
    \caption{Sketch of the domain considered in this paper: a rigid body immersed in a two-dimensional nematic liquid crystal. The liquid crystal is described by a director field $\bm{n} = (\cos\theta, \sin\theta, 0)$ with director angle $\theta(z)\in[0,\pi)$ for $z\in D$. The boundary of the body is shown as a solid curve, $\partial D$, with unit normal and tangent vectors $\bm{\n}(s)$ and $\bm{\s}(s)$, respectively. The effective boundary (used in the effective boundary technique, \S\ref{sec:moderatetangent}) is shown as a dashed curve, $\partial D_w$. } 
    \label{fig:sketch}
\end{figure}

Consider a two-dimensional nematic liquid crystal outside a simply connected body $D$ with boundary $\partial D$, as sketched in Fig.~\ref{fig:sketch}. Assuming the one-constant approximation, the director angle, $\theta(x,y)$, is described by the bulk free energy in \eqref{eq:1const_energy}. At the boundaries, we assume the Rapini--Papoular form of the surface anchoring energy, $\mathcal{F}_\mathrm{surface} \coloneqq W\sin^2(\theta-\phi)/2$, where $W$ is the anchoring strength and $\phi$ is the preferred orientation, which is generally a function of the local boundary orientation \citep{rp69}. For tangential anchoring, $\phi$ is the surface tangent angle. Combining the bulk and surface energies, the total energy of the configuration is defined as
\begin{equation}\label{eq:totalenergy}
    \mathcal{E} \coloneqq \frac{K}{2}\iint_D|\nabla \theta|^2\de A + \frac{W}{2}\int_{\partial D}\sin^2\left(\theta-\phi\right)\de s,
\end{equation}
where $\de A$ and $\de s$ are the infinitesimal surface area and arclength elements, respectively.

A variational principle applied to \eqref{eq:totalenergy} yields the equilibrium equations for the director angle, $\theta(x,y)$, (see App.~\ref{app:variational}):
\begin{equation}\label{eq:laplace}
    \nabla^2\theta = 0  \quad \text{in $D$},
\end{equation}
subject to the weak anchoring boundary condition,
\begin{equation}\label{eq:weak_anchor}
-K\frac{\partial \theta}{\partial\n} + \frac{W}{2}\sin\left[2(\theta-\phi)\right] =0 \quad \text{on $\partial D$},
\end{equation}
where $\bm{\n}\coloneqq-\bm{x}_s^\perp$ is the liquid crystal--pointing unit normal, as depicted in Fig.~\ref{fig:sketch}.

One of the goals of this work is to introduce a methodology for computing the force exerted on a body immersed in a liquid crystal.   The liquid crystal imposes stress according to the Ericksen stress tensor $\sigma\coloneqq -p I - K\nabla \bm{n}^T\nabla \bm{n} $ with hydrostatic pressure $p\coloneqq-K\lVert\nabla \bm{n}\rVert^2/2$ \citep{degennes1993,stewart2004}. On $\partial D$, there is a additional stress, which originates from the weak anchoring surface energy \citep{virga2018}. A virtual work argument yields the total surface traction acting on the boundary $\partial D$ (see App.~\ref{app:variational}):
\begin{equation}\label{eq:traction}
\begin{split}
    \bm{t}   &= K\left(\frac{1}{2}|\nabla\theta|^2\bm{\n}-\frac{\partial \theta}{\partial\n} \nabla\theta\right)+ \left(\mathcal{F}_\mathrm{surface}\bm{\s}-\frac{\partial \mathcal{F}_\mathrm{surface}}{\partial \phi} \bm{\n} \right)_{s}\\
    &= K\left(\frac{1}{2}|\nabla\theta|^2\bm{\n}-\frac{\partial \theta}{\partial\n} \nabla\theta\right)+ \frac{W}{2}\Big\{\sin(\theta-\phi)^2 \bm{\s}+\sin\left[2(\theta-\phi)\right]\bm{\n}  \Big\}_{s},
\end{split}
\end{equation}
where $\bm{\n}\coloneqq-\bm{x}_s^\perp$ and $\bm{\s}\coloneqq\bm{x}_s$ are the unit normal and tangent vectors, respectively, and subscript $s$ denotes an arc-length derivative, which is defined anti-clockwise. 

Overall, given a harmonic director field that satisfies \eqref{eq:weak_anchor}, the  energy and surface traction associated with the liquid crystal can be computed using \eqref{eq:totalenergy} and \eqref{eq:traction}, respectively. In this paper, we utilize a complex analysis formulation to derive analytical solutions to these equations, we shall introduce this formulation shortly.

\subsection{Dimensionless variables}

We non-dimensionalize all length scales with respect to a characteristic lengthscale associated with the immersed body, $a$, and introduce the dimensionless free energy and traction as $\hatE\coloneqq \mathcal{E}/K$ and $\hattaub\coloneqq a^2 \bm{t}/K$, respectively. The resulting equations are governed by a dimensionless anchoring strength $\w \coloneqq aW/K$. The dimensionless free energy of the liquid crystal, $\hatE$, may now be written as a boundary integral using the divergence theorem:
\begin{equation}\label{eq:totalenergy_dimensionless}
    \hatE = \frac{1}{2}\int_{\partial D}-\theta \frac{\partial \theta}{\partial\n}+w\sin^2\left(\theta-\phi\right) \de s.
\end{equation}
Henceforth, we shall only work in these dimensionless variables.

\subsection{Complex variables formulation}\label{sec:complex}

We introduce the complex coordinate $z\coloneqq x+\im y$ and define the complex director angle as
\begin{equation}
    \Omega(z) \coloneqq \tau - \im \theta,
\end{equation}
where $\tau(x,y)=\Re\Omega(z)$ is a harmonic conjugate of $\theta(x,y)=-\Im\Omega(z)$ (\ie~$\theta_x=\tau_y$ and $\theta_y=-\tau_x$). The gradient components of the director angle, $\nabla\theta=(\theta_x,\theta_y)$, are related to the complex angle above via 
\begin{equation}
    \theta_x - \im \theta_y = \im \Omega'(z).
\end{equation}

Since $\theta(x,y)$ is harmonic in $D$, $\theta_x-\im\theta_y$ must be  holomorphic in $D$ and $\Omega(z)$ is at least locally-holomorphic. In general, $\Omega$ may not be single-valued around the body, thus  the period  around  $\partial D$ must be defined, that is
\begin{equation}\label{eq:OmegaPeriod}
\oint_{\partial D}  \de \Omega \equiv \frac{1}{\im}\int_{\partial D}\theta_x -\im \theta_y \de z = \Upsilon-2\pi\im M,
\end{equation}
for some given real constant $\Upsilon$ and half-integer $M$. Note that $M$ corresponds to  the topological charge of the body, $\partial D$.

In these complex variables, the bulk-pointing normal vector, $\bm{\n}$, is represented as $y_s-\im x_s$ and the normal derivative of the director angle is written as $\theta_{\nu}=\Re[\Omega'(z)z_s]$. The boundary condition \eqref{eq:weak_anchor} is equivalent to a constraint on $\Omega(z)$,
\begin{equation}\label{eq:Omegabc}
\left(\big|\e^{\Omega(z)}\big|^2\right)_s+\w\Im\left[ \e^{2\im\phi}\e^{2\Omega(z)}\right]=0 \quad \text{on $\partial D$}.
\end{equation}

For example,  $\Omega(z)=-\im\alpha$ corresponds to a director angle $\theta(x,y)=\alpha$ uniform in space; the principal-value logarithm $\Omega(z) = -M\log(\e^{\im\alpha}z)$   corresponds to an isolated defect of topological charge $M$ located at $z=0$ and oriented with $\arg z= \alpha$; and $\Omega(z) = [\pi/(2\im\log R)-1]\log z$ corresponds to a nematic  contained in the annulus $1<|z|<R$ with strong normal and tangential anchoring on the inner  ($|z|=1$) and outer ($|z|=R$) boundaries, respectively. This last configuration is often called the `magical spiral' \citep{degennes1993} and has  $\Upsilon= \pi^2/\log R$ and $M=1$  in \eqref{eq:OmegaPeriod}.

The net free energy, \eqref{eq:totalenergy_dimensionless}, may be written in terms of $\Omega(z)$ as
\begin{equation}\label{eq:complexenergy}
\hatE = \frac{1}{4}\oint_{\partial D}\Im\left[ \left(\Omega(z)- \overline{\Omega( z)}\right) \Omega'(z)z_s\right]+ w \Re\left[1-\e^{\Omega(z)-\overline{\Omega( z)}}\e^{2\im\phi} \right]\de s,
\end{equation}
where the bar denotes a complex conjugate. Additionally, the surface traction $\hattaub\equiv(\hattau_x,\hattau_y)$ given by \eqref{eq:traction}
may be written as
\begin{equation}\label{eq:complextraction}
\hattau_x-\im \hattau_y=  \frac{1}{2\im}\Omega'(z)^2\frac{\partial  z}{\partial s} +\frac{w}{8}\left[\left(2+\e^{-2\im\phi}\e^{\overline{\Omega(z)}-\Omega(z)}-3\e^{2\im\phi}\e^{\Omega(z)-\overline{\Omega(z)}}\right)\frac{\partial \bar z}{\partial s}\right]_s.
\end{equation}

Overall, given a locally-holomorphic function, $\Omega(z)$, which satisfies the weak anchoring condition \eqref{eq:Omegabc} with a period \eqref{eq:OmegaPeriod}, the liquid crystal director angle, $\theta(x,y)=-\Im\Omega(z)$; free energy, \eqref{eq:complexenergy}; and surface traction, \eqref{eq:complextraction}, can all be determined.

\subsection{Net body force and torque}\label{sec:forcetorque}

The net dimensionless force, $(\hatF_x,\hatF_y)$, and torque, $\hatT$, acting on the  body are found by integrating the surface traction, $(\hat{t}_x,\hat{t}_y)$ given by \eqref{eq:complextraction}, and  surface moment, $x \hat{t}_y-y\hat{t}_x$, around $\partial D$, respectively. Integrating by parts, and imposing the boundary condition \eqref{eq:Omegabc} and the fact that $\exp[2\im(\theta-\phi)]\equiv \exp(\overline\Omega-\Omega-2\im\phi)$  is single-valued, results in the complex contour integrals 
\begin{subequations}\label{eq:complexforcetorque}
 \begin{gather}
 \hatF_x-\im \hatF_y  =\oint_{\partial D} \hat{t}_x-\im \hat{t}_y\de s = \frac{1}{2\im}\oint_{\partial D} \Omega'(z)^2 \de z, \\
\text{and} \quad \hatT   =\oint_{\partial D}x \hat{t}_y -y \hat{t}_x\de s =\frac{1}{2}\Re\left[\oint_{\partial D} z\Omega'(z)^2 \de z\right]+\Upsilon,
\end{gather}
\end{subequations}
 where  $\Upsilon$ is the period defined in \eqref{eq:OmegaPeriod} and   the centre of rotation  is assumed to be $z=0$, without loss of generality.

Assuming there are no defects within the liquid crystal, \ie~$\Omega(z)$ is holomorphic in $D$, we are free to deform the integration contours in \eqref{eq:complexforcetorque} provided they remain outside the body. For a bounded director angle, $\Upsilon = 0$ and $\Omega'(z) \sim -M/z$ as $|z|\to \infty$, for some half-integer $M$ corresponding to the topological charge of the body, as introduced in \eqref{eq:OmegaPeriod}. Thus, by taking the contours in \eqref{eq:complexforcetorque} to infinity, we immediately find that there is no net force or torque acting on a immersed connected body when the director angle is bounded (\ie~$\hatF_x=\hatF_y=\hatT=0$). Note that a non-zero torque ($\hatT\neq 0$) induces a logarithmically growing director angle, that is  $\theta(z)\equiv-\Im\Omega(z)\sim M\arg z -\hatT\log|z|/[2\pi(1-M)]$ as $|z|\to \infty$ --- \ie~$\Upsilon=\hatT/(1-M)$ in \eqref{eq:OmegaPeriod}. 

\emph{Analogy with d'Alembert's paradox.} A classical and paradoxical result in potential fluid flow states that an inviscid fluid presents no resistance to a body which moves through it with steady translational motion \citep{Batchelor67}. The resolution of this paradox is found with the incorporation of a viscous boundary layer on the body, which promotes the separation of vorticity from the surface in its wake. Here, we find an analogous result: there is no net force or torque acting on a immersed connected body when the director angle is bounded (\ie~$\hatF_x=\hatF_y=\hatT=0$). Unlike in potential flow theory, however, this result does not challenge our physical experience, and thus we do not seek further resolution in this setting.

\emph{Analogy with Stokes' paradox.} Another classical fluid mechanical paradox appears in Stokes flow. Namely that the flow due to a cylinder upon which there is a non-zero force is logarithmic in the distance from the body, and so the fluid velocity in an infinite domain is unbounded \citep{Batchelor67}. The resolution of this apparent paradox is found by the reintroduction of inertial effects (\eg~via the Oseen equations). The analogous result here is that the director angle far from a body immersed in a LC with a non-zero torque is, similarly, logarithmic, and thus the associated elastic energy is unbounded. The question of solving for the equilibrium director field around a two-dimensional body with a non-zero torque is therefore similarly ill-posed. Reintroducing LC dynamics would regularize this singular behavior by presenting a finite speed of propagation of information from near the body outward towards infinity.

A non-zero net force or torque on a body at equilibrium requires either a secondary body or the existence of  a defect within the liquid crystal. For  example,  consider the complex director angle $\Omega(z)= M\log[(z-\epsilon)/(z+\epsilon)]$  for a half-integer $M$. This corresponds to an unconfined director field with two defects at $z=\pm\epsilon$ of topological charge $\mp M$, respectively. The net force acting on one of these defects can be computed by integrating  $\Omega'(z)^2= 4M^2\epsilon^2/(z^2-\epsilon^2)^2$ around a contour only containing said defect, \ie~\myeqref{eq:complexforcetorque}{a}.  Cauchy's residue theorem then yields the forces $(F_x^{\pm},F_y^\pm) = (\pm \pi M^2/\epsilon,0)$  for the  defects of  charge $\pm M$, respectively. We, hence, find that the defects are attracted to each other with a force proportional to the inverse of the separation distance and the square of their topological charge \citep{degennes1993,gsv02,hs20}, though their mobilities and migration speeds in a dynamic setting can differ substantially \citep{tdy02,ts19}.

\section{Effective boundary technique for finite anchoring strength}\label{sec:moderatetangent}

While  infinite anchoring ($w=\infty$) is mathematically appealing, real anchoring energies are finite, so we now turn to the case of large, but finite, anchoring strengths. As $w\to \infty$, the strong  anchoring boundary condition is recovered from \eqref{eq:Omegabc}, \ie
\begin{equation}\label{eq:Stronganchoring}
\Im \e^{\Omega(z)+\im\phi}=\Oh(1/w) \quad \text{on $\partial D$}.
\end{equation} 
This  boundary condition is invariant under a conformal map, thus the corresponding problem can be solved using a wide-range of complex analysis tools \citep{af03}. However, solutions to these Dirichlet problems are known to require singularities at  boundaries, which correspond to  `surface defects' within the liquid crystal. This is problematic as defects give rise to an infinite free energy. This issue is often resolved by introducing a `melting region' local to the defects \citep{degennes1993}, or by studying a more general alignment tensor theory in which this melting from a nematic to isotropic phase is tracked by a Maier-Saupe scalar order parameter~\citep{Hess75,skh95,sl03}. Here, we instead seek a correction to the strong anchoring constraint as $\w\to \infty$.

For convenience, we denote $\Phi(s)\coloneqq \e^{\Omega(z)+\im\phi}$, which is defined for $z(s)\in\partial D$. The boundary condition, \eqref{eq:Omegabc}, can then  be written as the nonlinear constraint
\begin{equation}\label{eq:stronganchoring_Phi}
\Re \Phi(s)\Re \Phi'(s)+\Im \Phi(s)\Im \Phi'(s)+w\Im \Phi(s)\Re \Phi(s)=0.
\end{equation}
Using the fact that $\Im \Phi = \Oh(1/w)$ as $w\to \infty$, \ie~\eqref{eq:Stronganchoring}, we immediately find that
\begin{equation}\label{eq:moderateanchoring}
\Im\Phi(s)+\frac{1}{w}\Re\Phi'(s)=\Oh(1/w^3),
\end{equation}
as $w\to \infty$. 

The linear constraint \eqref{eq:moderateanchoring} contains the desired correction to the strong anchoring condition when the anchoring strength is large, but finite. This constraint is analogous to a Robin boundary condition for $\Im \Phi$; however, unlike its Dirichlet counterpart (when $w=\infty$), it is not preserved under a conformal map and, thus, the most appealing tools of complex analysis do not immediately apply. 

It is known, however, that the configuration of a nematic with weak anchoring (finite $w$) near a flat boundary is equivalent to that of a nematic with strong anchoring ($w=\infty$) on a different surface, internal to the boundary, which is recessed from the surface by the extrapolation length $K/W\equiv a/w$ \citep{degennes1993,rp69}. Below, we use this idea to show that imposing the finite anchoring constraint \eqref{eq:moderateanchoring} on $\partial D$ is mathematically equivalent to imposing the strong anchoring constraint \eqref{eq:Stronganchoring} on some `effective domain' boundary, $\partial D_w$, which is the physical  boundary moved by the extrapolation length. The effective boundary, $\partial D_w$, is sketched in Fig.~\ref{fig:sketch}.

The above result follows from noting that \eqref{eq:moderateanchoring} can  be written as 
\begin{equation}
\Im\Phi(s+\im/w)=\Oh(1/w^3),
\end{equation}
where we define  $\Phi(s+\im/w)\coloneqq \Phi(s)+\Phi'(s)\im/w- \Phi''(s)/(2w^2) +\Oh\left(1/w^3\right)$. Substituting in $\Phi=\e^{\Omega(z)+\im\phi}$ then yields the boundary condition
\begin{equation}\label{eq:modstronganchoring}
    \Im \e^{\Omega(z)+\im\tilde{\phi}} =  \Oh(1/w^3)\quad \text{on $\partial D_{\w}$},
\end{equation}
 where $\tilde{\phi}(s)\coloneqq\phi(s)-\phi''(s)/(2w^2) + \Oh(1/w^3)$  and $\partial D_\w$ is the boundary $\partial D$  displaced by $-\bm{\n}(s)/\w-\bm{\s}'(s)/(2\w^2) +\Oh(1/w^3)$ for the unit normal and tangent vectors  $\bm{\n}(s)=-\bm{x}'(s)^\perp$ and  $\bm{\s}(s)=\bm{x}'(s)$, respectively  --- \ie~$z(s)\in \partial D$ is mapped to $z(s+\im/w)\in\partial D_w$.

Above, we have implicitly assumed that the boundary curve, $\partial D$, is non-singular; however, domains with corners are ubiquitous in the literature \citep{lms09,dm12,Muvsevivc17}. A local asymptotic analysis shows that mapping a corner in $\partial D$ to a similar corner  in $\partial D_w$ is accurate up to $\Oh(1/w^3)$ as $w\to \infty$ --- \ie~the same order of accuracy as \eqref{eq:modstronganchoring}.

Overall, we have shown that finding a locally-holomorphic function in $D$  that satisfies \eqref{eq:Omegabc}  on $\partial D$ is mathematically equivalent  to finding a locally-holomorphic function in $D_w$ that satisfies \eqref{eq:modstronganchoring} on $\partial D_w$, up to $\Oh(1/w^3)$  as $\w\to\infty$. Here, we have analytically continued $\Omega(z)$ inside the narrow region $D_\w\setminus D$. It should  be noted that the modified anchoring angle in \eqref{eq:modstronganchoring}, \ie~$\tilde\phi(s)\sim \phi(s)-\phi''(s)/(2w^2)$,  may differ from the original anchoring angle in \eqref{eq:Omegabc}, \ie~$\phi(s)$. The equivalence of these two problems shows that any  defects on the immersed boundary for $w=\infty$ are located inside the boundary for finite $w$, thus regularizing the energy. Furthermore, it provides a method for solving the finite anchoring problem on a domain $D$: first one constructs the effective domain, $D_w$, via displacing  $\partial D$ by $-\bm{\n}(s)/w-\bm{\s}'(s)/(2w^2) +\Oh(1/w^3)$\footnote{As a first approximation, one can instead  move the boundary a distance $1/w$ in the normal direction, however the  solution will then only be  accurate  to  $\Oh(1/w^2)$ as  $w\to \infty$. In some cases, higher-order terms are required for computing the net energy, \eqref{eq:complexenergy}, and force, \eqref{eq:complexforcetorque}.}, whilst ensuring any corner angles are preserved; then one  solves the strongly anchored problem on $D_w$, for which the standard techniques of complex analysis, in particular conformal mapping, may be used.

\subsection{Tangential (homogeneous) anchoring and complex potentials}\label{sec:potential}

Although the results presented thus far hold for any anchoring angle, $\phi$, we are particularly interested in the case of tangential (homogeneous) anchoring with $\phi$ being  tangent to $\partial D$. For tangential anchoring,    $\phi(s) \coloneqq \arg z'(s)\mod\pi$ for  $z(s)\in\partial D$. Furthermore, since
\begin{equation}
    \arg\left[z'(s)+\frac{\im}{w}z''(s)-\frac{1}{2w^2}z'''(s)\right]= \phi(s)-\frac{1}{2w^2}\phi''(s)+\Oh(1/w^3)\mod\pi,
\end{equation}
as $w\to \infty$,  the modified anchoring angle, $\tilde\phi(s)\sim \phi(s)-\phi''(s)/(2w^2)$, is  tangent  to the effective domain, \ie~$\tilde\phi(s) \coloneqq \arg \tilde{z}'(s) \mod\pi$ for  $\tilde{z}(s)\in\partial D_w$. With this, the effective boundary technique (derived in \S\ref{sec:moderatetangent}) says that imposing weak tangential anchoring on $\partial D$ is equivalent to imposing strong tangential anchoring on $\partial D_w$ up to $\Oh(1/w^3)$ as $w\to \infty$. (This equivalency also holds for normal, homeotropic, anchoring with $\phi(s)=\pi/2+\arg z'(s) \mod \pi$ for  $z(s)\in\partial D$ or $\partial D_w$. Moreover, since a homeotropic director field is perpendicular to its homogeneous counterpart, one can transform a  homeotropic anchoring problem into a homogeneous anchoring problem by taking $\Omega_\mathrm{homeo}(z) =  \Omega_\mathrm{homo}(z)+\im\pi/2$.)

Strong tangential anchoring is reminiscent of a vanishing boundary-flux  problem in potential theory with   $\Omega(z)$  corresponding to the logarithm of the derivative of a complex potential $f(z)$, \ie~$\Omega(z) = \log f'(z)$ \citep{af03,Acheson1990}; for example, $f(z)=z \e^{-\im \alpha}$ corresponds to a director angle $\theta=\alpha$ uniform in space. However,  the problem here differs from  the typical potential problem since $\Omega(z)$ is not necessarily single-valued due to the periods in \eqref{eq:OmegaPeriod}.

Inspired by the above discussion, it is natural to remove the multivaluedness of $\Omega(z)$ then introduce an analogous complex potential. That is, we write
\begin{equation}
    \Omega(z) = \log f'(z) + g(z),
\end{equation}
where $g(z)$ is a locally-holomorphic function, which accounts for the period in \eqref{eq:OmegaPeriod}, and $f'(z)$ is a holomorphic function, which satisfies the boundary condition  on $\partial D$. 

If we also restrict our attention to the case when $M=0$ in \eqref{eq:OmegaPeriod} (\ie~the body, $\partial D$, has a vanishing topological charge), we then have enough freedom to choose $g(z)$ such that
\begin{equation}\label{eq:upsilonBC}
\Im g=0\quad \text{on $\partial D_w$.}
\end{equation}
The boundary condition \eqref{eq:modstronganchoring} can then be integrated once to obtain 
\begin{equation}\label{eq:fBC}
    \Im f= \Oh\left(1/w^3\right)\quad \text{on $\partial D_{\w}$},
\end{equation}
where  we have fixed the gauge of $f$ such that the integration constant vanishes. However, the boundary condition \eqref{eq:fBC} does not necessarily yield a unique solution since, although $f'(z)$ is single-valued, its primitive may be multivalued. Uniqueness is restored by specifying the period of $f(z)$ around $\partial D$, \ie
\begin{equation}\label{eq:fCirc}
    \oint_{\partial D} \de f =\Gamma,
\end{equation}
for some real constant $\Gamma$. This period is analogous to specifying a constant circulation around a body in potential theory \citep{Acheson1990,af03,vn06,Crowdy20}, and it affects the locations of topological defects on the surface, as we shall see. 

Overall, when $M= 0$, the homogeneous anchoring problem comes down to finding two locally holomorphic potentials, $g(z)$ and $f(z)$, which satisfy the boundary conditions \eqref{eq:upsilonBC} and \eqref{eq:fBC}  with given periods \eqref{eq:OmegaPeriod} and \eqref{eq:fCirc}, respectively. Although this formulation may appear daunting, $g(z)$ and $f(z)$ are both equivalent to a complex potential in a typical potential theory problem \citep{Acheson1990,af03,Crowdy20}. We can, thus, utilize the vast number of results which already exist within the literature.

\section{Example 1: Immersed cylinder with tangential anchoring}\label{sec:ex1}

\begin{figure}
    \centering
    \includegraphics[width=0.8\textwidth]{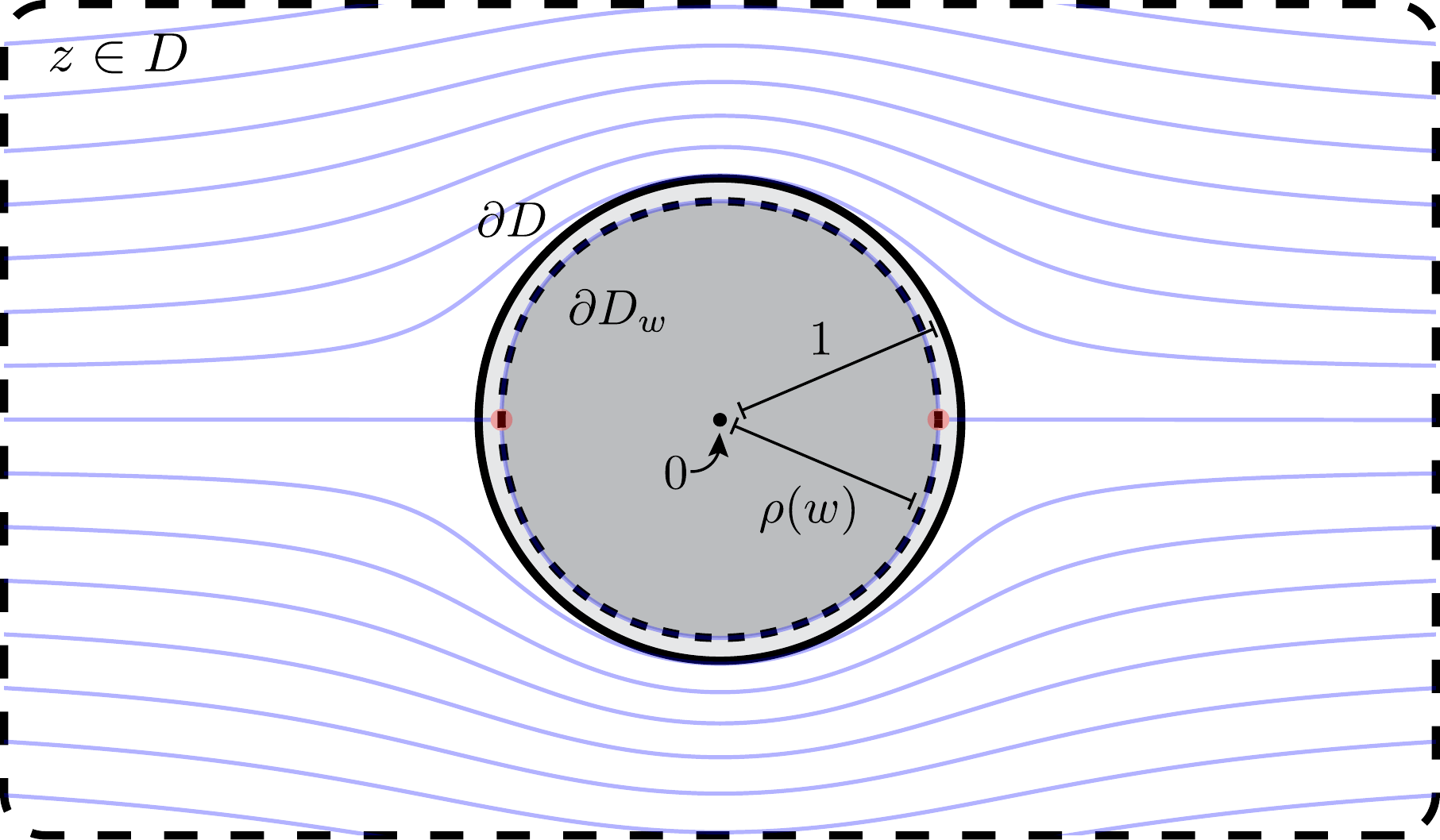}
    \caption{\emph{Example 1.} Two-dimensional liquid crystal outside a unit cylinder ($\partial D$, black solid curve) subject to finite tangential anchoring. The effective domain boundary is a shrunken cylinder of radius $\rho(w)$ ($\partial D_w$, black dotted curve). The director field inside the effective domain, \eqref{eq:ex1_theta}, is shown as faded blue curves for $w=10$ and period which minimizes the free energy, $\Gamma=0$.}
    \label{fig:ex1_sketch}
\end{figure}

Consider a liquid crystal outside  a cylinder with dimensionless unit radius, we denote this region $D$.  The liquid crystal is assumed to be oriented  with the $x$-axis in the far-field, thus $\Omega(z)\to 0$ as $|z|\to \infty$, and subject to finite tangential (homogeneous) anchoring on the cylinder, \ie~$\Omega(z)$ satisfies \eqref{eq:moderateanchoring} with $\phi=\arg z'(s)\mod \pi$ on $|z|=1$. (Although we have assumed finite anchoring  with $w\gg 1$, the solution derived in this section ultimately satisfies the weak  anchoring condition, \eqref{eq:Omegabc}, for any $w\geq0$.)  This configuration is plotted in Fig.~\ref{fig:ex1_sketch}.

Since $\Omega(z)$ is holomorphic in $D$, the contour in the  period-defining integral \eqref{eq:OmegaPeriod} can be deformed freely within the liquid crystal. Sending this contour off to infinity and imposing the far-field conditions, it follows that $\Omega(z)$ must be single-valued with the periods in \eqref{eq:OmegaPeriod} vanishing  (\ie~$\Upsilon=M=0$). Following \S\ref{sec:potential}, we  introduce a locally holomorphic potential, $f(z)$,  defined such that $\Omega(z)=\log f'(z)$ and with its period around the cylinder as yet free, \ie~$\Gamma$ in \eqref{eq:fCirc}. (Note that $g(z)$ is omitted here since $\Omega(z)$ is already single-valued.) 

The first step of the effective boundary technique  is to construct the effective domain, $D_w$. Since the normal and tangent vectors  of the cylinder are $\bm{\n}(s)=(\cos s,\sin s)$ and $\bm{\s}(s)=(-\sin s,\cos s)$, respectively, where $s$ is the complex argument, the unit cylinder is displaced by $-1/\w+1/(2\w^2)$ in the normal direction. Thus, the effective domain, $D_w$, is the region outside a cylinder of dimensionless radius $\rho(w)\coloneqq 1-1/\w+1/(2\w^2)+\Oh(1/\w^3)$ --- as shown in Fig.~\ref{fig:ex2_sketch}. Analytically continuing $f(z)$ inside the annulus $\rho(w)<|z|<1$ then yields the equivalent problem: find an $f(z)$ such that
\begin{subequations}\label{eq:ex1_problem}
\begin{align}
f(z) \text{ locally holomorphic} &\qquad \text{ in $D_w$,}\\
 \Im f(z)=0 &\qquad \text{ on $\partial D_w$,}\\
 f(z)\sim  z &\qquad \text{ as $|z|\to\infty$,}
\end{align}
\end{subequations}
and with period $\oint_{D_w}\de f = \Gamma$. 

The potential problem \eqref{eq:ex1_problem} is a historic problem in the  field of fluid dynamics, corresponding to the potential flow around a circular cylinder. Its solution can be found in many textbooks \citep{af03,Acheson1990}: 
\begin{equation}\label{eq:ex1_sol}
    f(z) = z+ \frac{\rho^2}{z}+\frac{\Gamma}{2\pi\im}\log\frac{z}{\rho}.
\end{equation}
This potential corresponds to the director angle
\begin{equation}\label{eq:ex1_theta}
\theta(x,y)\equiv-\arg f'(z) = 2\arg z-\arg\left(z-\rho \e^{\im\gamma}\right)-\arg\left(z+\rho \e^{-\im\gamma}\right),
\end{equation}
from which it is found that the defect locations are set by the period, $\Gamma$, through the simple relation $\sin\gamma\coloneqq\Gamma/(4\pi\rho)$. In fact, this solution not only satisfies the finite anchoring condition \eqref{eq:moderateanchoring} for $w\gg 1$,  but also satisfies the weak anchoring condition \eqref{eq:stronganchoring_Phi} for any $w\geq 0$, provided $\Gamma=0$ and we set $\rho(w)\coloneqq({\sqrt{1+4/\w^2}-2/\w})^{1/2}$. We assume this choice of effective radius henceforth. 

The solution \eqref{eq:ex1_theta}  has  previously been derived by \cite{burylov1994} using a superposition of separable solutions to Laplace's equation. Furthermore, it  is equivalent to an unrestricted director field with three  defects points: two $-1$ defects at $z= \rho \e^{\im\gamma}$ and $z=- \rho \e^{-\im\gamma}$ and a $+2$ defect at $z=0$. Note that these defects do not reside in the bulk liquid crystal domain for finite $\w$; however, the $-1$ defects do correspond to surface defects as  $\rho(w\to\infty)\to 1$.

The period $\Gamma$, or equivalently the positions of the $-1$ defects, is chosen to minimize the energy of the static liquid crystal, \eqref{eq:complexenergy}. This energy reduces to the complex contour integral
\begin{equation}\label{eq:ex1_energycontour}
\hatE = \frac{1}{4}\Im\left[2\pi\im w+\oint_{|z|=1} w\frac{z f'(z)}{\bar{f}'(1/z)} -\log\bar{f}'(1/z) \frac{f''(z)}{f'(z)}\de z\right],
\end{equation}
after imposing the  Schwarz function of the cylinder, $\bar z =1/z$, \citep{af03,Shapiro1992}. Inserting \eqref{eq:ex1_sol} into \eqref{eq:ex1_energycontour} results in a contour integral, which can be evaluated using Cauchy's residue theorem via accounting for the  poles at the three defect points: $z=0$, $\rho \e^{\im\gamma}$, and $-\rho \e^{-\im\gamma}$; this yields
\begin{equation}\label{eq:ex1_AnalyticEnergy}
\hatE = \frac{\pi}{2}w(1-\rho(w)^2)-\pi\log\left|1-\rho(w)^2\right|-\pi\log\left|1+\rho(w)^2 \e^{2\im\gamma}\right|.
\end{equation}
Further details  on evaluating \eqref{eq:ex1_energycontour} can be found in App.~\ref{app:energy_ex1}.

It is evident from  \eqref{eq:ex1_AnalyticEnergy} that $\gamma=0$ minimizes the net free energy. This result is not surprising due to the up--down symmetry of the problem; however, this is not necessarily the case  for other geometries (see \S\ref{sec:ex2}, for example). We also note that $\hatE=\pi+\pi \log(w/4) - \pi/2\log|1-\Gamma/(4\pi)|+ \Oh(1/w)$ as $w\to \infty$, \ie~the energy is logarithmically singular in the strong anchoring limit (a consequence of the surface defects at $z\sim \pm \e^{\pm\im\gamma}$).

Above, we assumed the immersed cylinder only locally disturbs the liquid crystal, so that the director field is oriented at a constant angle in the far-field. Another physical, but energetically unfavourable, configuration is that the cylinder appears as a point defect of topological charge $M\neq 0$ when viewed from afar, as introduced in \eqref{eq:OmegaPeriod}. This would correspond to imposing  $\Omega(z)\sim -M\log z$ as $|z|\to \infty$. In this case, one can show that the resulting director angle is
\begin{equation}
\theta(z)= (2-M)\arg z-\arg\left(z^{2(1-M)}-\rho_M^{2(1-M)}\right),
\end{equation}
 with  effective cylinder radius, $\rho_M(w)$, given by   
\begin{equation}
\big[\rho_M(w)\big]^{2-2M} \coloneqq \sqrt{1+\frac{4(1-M)^2}{\w^2}}-\frac{2(1-M)}{\w},
\end{equation}
for any $w\geq0$. This solution corresponds to $3-2M$ defects: $2(1-M)$ defects of strength $-1$ on  $|z|=\rho_M(w)$ and a  defect of strength $2-M$ at $z=0$. 

We finish this example by noting that, since the potential problem \eqref{eq:ex1_problem} is invariant under a conformal map, the  director angle \eqref{eq:ex1_sol} describes the director field outside \emph{any}  connected body,  provided the conformal map from  the wanted extended domain, $\zeta\in D_\w$, to the punctured domain, $|z|\geq \rho(w)$, can be obtained. Although such a conformal map must exist by the Riemann mapping theorem, determining such a mapping is notoriously difficult even for simple domains \citep{af03}. Furthermore, although the effective domain, $\zeta\in D_\w$, is  mapped onto $|z|\geq \rho(w)$, the original domain, $\zeta\in D$, may not  be mapped onto $|z|\geq 1$ since size and curvature are not necessarily preserved under a conformal map. To demonstrate this, we now consider a slightly more complicated domain: the region outside an equilateral triangle.

\section{Example 2: Immersed triangle with tangential anchoring}\label{sec:ex2}

\begin{figure}
    \centering
    \includegraphics[width=0.8\textwidth]{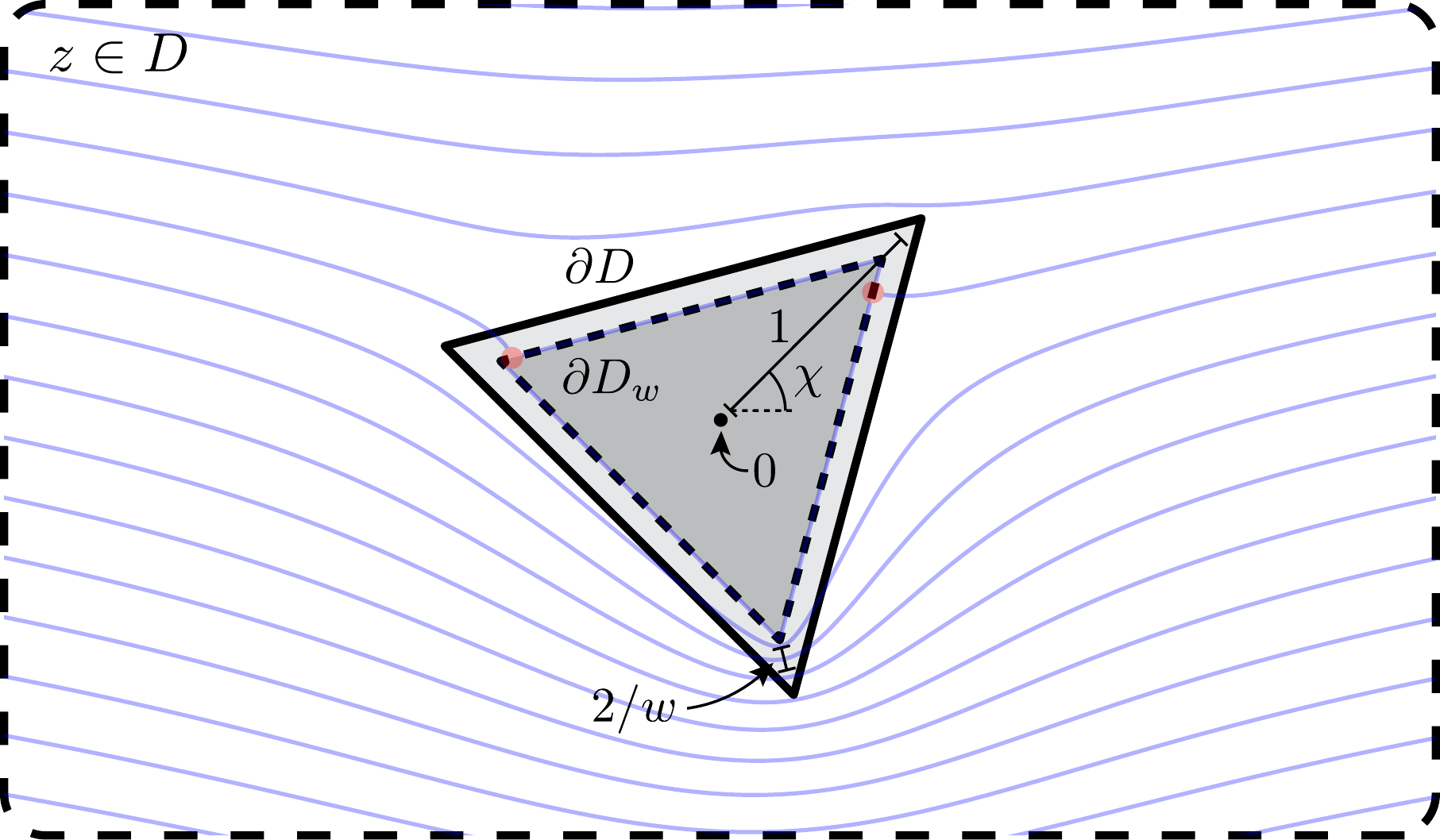}
    \caption{\emph{Example 2.} Two-dimensional liquid crystal outside a triangle with corners at the roots of $z^3=\e^{3\im\chi}$ ($\partial D$, black solid curve) subject to weak (finite) tangential anchoring. The effective domain boundary is a similar triangle with corners at the roots of $z^3=(1-2/w)^3\e^{3\im\chi}$  ($\partial D_w$, black dotted curve). The director field inside the effective domain,  \eqref{eq:ex2_theta}, is shown as faded blue curves for $w=10$, $\chi=\pi/4$, and  period  which minimizes the free energy, $\Gamma\approx 3.14$.}
    \label{fig:ex2_sketch}
\end{figure}

We next consider an immersed equilateral triangle with vertices at the roots of $z^3=\e^{3\im\chi}$. As before, the liquid crystal is assumed to be oriented with the $x$-axis in the far-field, \ie~$\Omega(z)\to 0$ as $|z|\to\infty$, and subject to finite tangential (homogeneous) anchoring on the triangle, \ie~$\Omega(z)$ satisfies \eqref{eq:moderateanchoring} with $\phi=\arg z'(s)\mod \pi$ on $\partial D$. Assuming there are no  defects within the liquid crystal,  the complex director angle, $\Omega$, must be single-valued. We may, thus, introduce a complex potential, $f(z)$, defined such that $\Omega(z)=\log f'(z)$  with period $\Gamma$, as in \eqref{eq:fCirc}.  This configuration is plotted in Fig.~\ref{fig:ex2_sketch}.

For large anchoring strengths ($w\gg1$), the effective boundary technique is of use (\S\ref{sec:moderatetangent}). The first step is to construct the effective domain, $D_w$.  As the tangent vectors  are constant, the triangle edges are displaced by $1/w+\Oh(1/w^3)$ in the normal direction with the corner angles preserved. Consequently, we find that the effective domain, $D_w$, is the region outside a triangle with corners at the roots of  $z^3=(1-2/w)^3\e^{3\im\chi}$, as shown in  Fig.~\ref{fig:ex2_sketch}.  Analytically continuing $f(z)$ into $D_w\setminus D$ yields the same potential problem 
 as \eqref{eq:ex1_problem}, but with $D_w$ now denoting the domain outside the effective triangle.

Since the potential problem, \eqref{eq:ex1_problem}, is invariant under a conformal map, the solution  \eqref{eq:ex1_sol} can be used here,  provided a conformal map from the triangular domain, $z\in D_w$, to the punctured domain, $|\zeta|\geq 1$, can be found. Such a conformal map can be constructed using a Schwarz--Christoffel mapping, which maps the unit disk (or upper half-plane) onto a simple polygon \citep{af03,dt02}.

By the Riemann-mapping theorem, there are  three real degrees of freedom when constructing a conformal map  \citep{af03}. We shall fix the pre-images of the effective triangle corners (\ie~the roots of  $z^3=(1-2/w)^3\e^{3\im\chi}$) to be the roots of $\zeta^3=\e^{3\im\chi}$.  The corresponding Schwarz--Christoffel mapping is
\begin{equation}\label{eq:ex2:SchwarzChristofell}
    z(\zeta) = A+B\int^\zeta \left(s^3-\e^{3\im\chi}\right)^{2/3}\frac{\de s}{s^2},
\end{equation}
for complex constants $A$ and $B$, which are fixed by ensuring that the corners of the triangle are mapped to their respective  pre-images. After evaluating the constants, the mapping \eqref{eq:ex2:SchwarzChristofell} can be written as
\begin{equation}\label{eq:ex2_zzeta}
    z(\zeta) = \left(1-\frac{2}{w}\right)\frac{h(\e^{3\im\chi}/\zeta^{3})}{h(1)}\zeta,
\end{equation}
for the hypergeometric function $h(\zeta)\coloneqq {}_2 F_1 (-2/3,-1/3;2/3;\zeta)$ \citep[as defined by][for example]{Abramowitz1964}.

In the punctured domain, $|\zeta|\geq 1$, the  locally-holomorphic function $F(\zeta)\coloneqq f(z(\zeta))$  satisfies $\Im F(\zeta)=0$ on $|\zeta|=1$, $F(\zeta)\sim (1-2/w)\zeta/h(1)$  as $|\zeta|\to\infty$, and has a period $\Gamma$ around $|\zeta|=1$. This problem is equivalent to \eqref{eq:ex1_problem} and, thus, has the same solution form:
\begin{equation}\label{eq:ex2_Fzeta}
    F(\zeta) =\frac{1-2/w}{h(1)}\left( \zeta+\frac{1}{\zeta}\right) +\frac{\Gamma}{2\pi \im}\log \zeta.
\end{equation}
Taking the derivative of \eqref{eq:ex2_Fzeta} with respect to $z$ yields the  director angle
\begin{equation}\label{eq:ex2_theta}
    \theta(z)\equiv-\arg f'(z)  =\frac{2}{3}\arg\left[\zeta(z)^3-\e^{3\im\chi}\right]-\arg\left[\zeta(z)-\e^{\im\gamma}\right]-\arg\left[\zeta(z)+\e^{-\im\gamma}\right],
\end{equation}
for $\sin\gamma\coloneqq h(1)\Gamma/[4\pi(1-2/w)]$ and $\zeta(z)$ given by the inverse of \eqref{eq:ex2_zzeta}.

As in Example 1 (\S\ref{sec:ex1}), we will determine the period, $\Gamma$, so as to minimize the net free energy of the liquid crystal,  \eqref{eq:complexenergy}. However, unlike the cylindrical case, the free energy can not be written as a closed integral of an analytic function  since a triangle does not have an analytic Schwarz function, $\bar z = S(z)$, and the boundary angle, $\arg z_s$, is only defined piecewise \citep{af03,Shapiro1992}. Instead we separate \eqref{eq:complexenergy} into three finite contour integrals, corresponding to the three edges of the triangle. These contour integrals are then evaluated numerically for given $\Gamma$, $w$, and $\chi$. The resulting energy is then minimized numerically to determine $\Gamma=\Gamma_{\min}(w,\chi)$ --- we use \texttt{fminsearch} with numerical quadrature in  \textsc{Matlab}. (Further details on evaluating the energy can be found in App.~\ref{app:energy_ex2}.)

\begin{figure}
    \centering
    \includegraphics[width=0.9\textwidth]{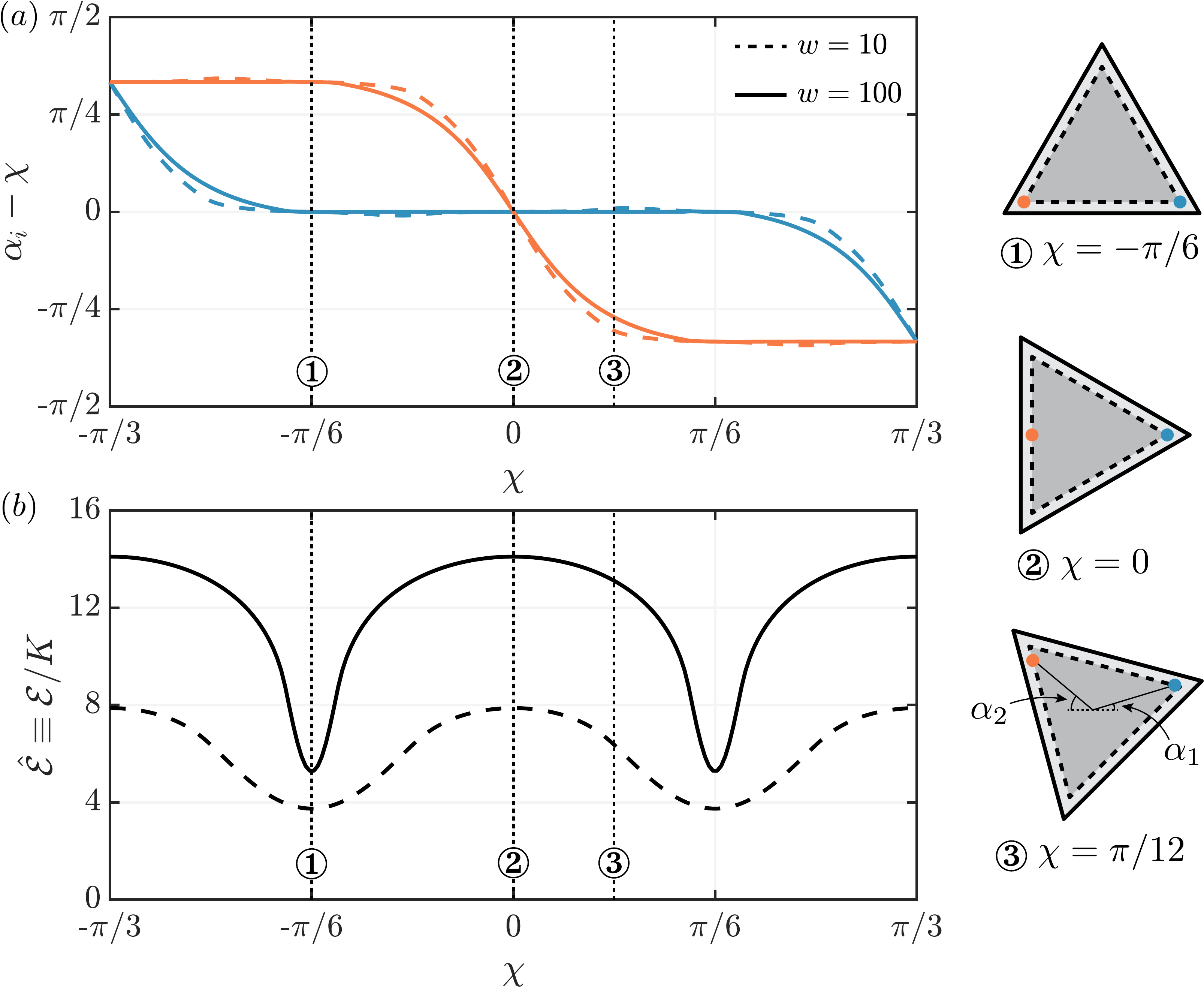}
    \caption{\emph{Example 2.} (a) Plot of the complex arguments of the two $-1$ defects as a function of the triangle orientation, $\chi$, for $w=10$ (dashed curve) and $w=100$ (solid curve).  The rightmost defect is described by the blue curve with $\alpha_1 \coloneqq \arg\left[z(\e^{\im\gamma_{\min}})\right]$, whilst the leftmost defect is described by the red curve with $\alpha_2 \coloneqq  \arg\left[-z(-\e^{-\im\gamma_{\min}})\right]$. Here,  $\Gamma=\Gamma_{\min} \equiv 4\pi(1-2/w)\sin \gamma_{\min} /h(1)$ is the period which minimizes the net free energy, \eqref{eq:complexenergy}. In \ding{172}--\ding{174}, the positions of the $-1$  defects are plotted as coloured dots for $w=10$ and the  labelled $\chi$. Note that  at least one of the defects lies approximately at a corner of the effective triangle for $w=100$ (\ie~$\alpha_1-\chi\approx 0$ or $\alpha_2-\chi\approx\pm\pi/3$), however this is not the case for $w=10$ (as seen in \ding{174}, for example). (b) Plot of the  net free energy as a function of $\chi$, corresponding to   the solutions presented in (a). For both $w=10$ (dashed curve) and $w=100$ (solid curve), it is evident that the free energy is smallest when the triangle is  pointing upwards  ($\chi=-\pi/6$) or downwards ($\chi=\pi/6$).}
    \label{fig:TriangleDislocation}
\end{figure}

It is evident from \eqref{eq:ex2_theta} that there are five defect points on the effective triangle:  three  defects at the corners, \ie~at the roots of $z^3=(1-2/w)^3\e^{3\im\chi}$, and two $-1$ defects at $z=z(\pm \e^{\pm\im\gamma})$. Although the locations of the corner defects are fixed for any given $w$ and $\chi$, the $-1$ defects are dependent on $\Gamma=\Gamma_{\min}(w,\chi)$, thus their locations are not immediately apparent. As $\w\to\infty$, we  observe that at least one of the $-1$ defects lies at a corner of the effective triangle (this can be seen approximately for $w=100$ in Fig.~\myref{fig:TriangleDislocation}{a}, for example).  This can be understood by noting  that most of the liquid crystal deformation occurs in the vicinity of defects, thus the energy is minimized by minimizing the number of topological defects. For smaller anchoring strengths, however, the defects do not lie in the physical domain; thus, placing one of the $-1$ defects at a corner of the effective triangle is not necessarily the energy-minimizing solution (this can be seen for $w=10$ in Fig.~\myrefnb{fig:TriangleDislocation}{\ding{174}}, for example).

\emph{Analogy with the Kutta condition.} Here we observe another analogy with the classical theory of potential  flow  past a body, in particular the placement of the rear stagnation point at a sharp trailing edge. The selection of the circulation which aligns these two is known as the Kutta condition, and the lift which results in that setting is known as the Kutta--Joukowski lift theorem \citep{Acheson1990}. We observe a very similar structure in the case of strong anchoring ($w\to \infty$); however, there is no direction associated with a nematic, thus  either defect could reside at a corner (there is no `trailing' edge). Moreover, since the force and torque  on the body at equilibrium must be zero (due to the analogue with  Stokes' paradox, as described in \S\ref{sec:forcetorque}), here there is no `lift', regardless of the circulation/period.

For each orientation of the triangle, $\chi$, we are able to minimize the net free energy over the period $\Gamma$; thus,   each $\chi$  has an  energetic cost associated with it. In Fig.~\myref{fig:TriangleDislocation}{b}, this minimized free energy is plotted as a function of $\chi$ for $w=10$ and $w=100$. It is apparent that that this  energy is smallest  when the triangle is either pointed upwards ($\chi=-\pi/6$) or downwards ($\chi=\pi/6$): one of the sides of the triangle is aligned with the relaxed orientation of the liquid crystal, \ie~the $x$-axis. This observation appears to hold for all values of the anchoring strength, $w$. Furthermore, for these values of $\chi$, the two $-1$ defects appoximately lie at the two vertices associated with the horizontal side of the effective triangle (see Fig.~\myrefnb{fig:TriangleDislocation}{\ding{172}}, for example). This is consistent with the experimental observations of \cite{lms09}, who observed that regular polygonal colloids reorient themselves until one of their sides aligns with the relaxed direction of the liquid crystal. The analysis of this phenomenon requires the introduction of  dynamics within the liquid crystal and, thus, is left for future work.

\section{Example 3: Mobility of a Janus particle with hybrid anchoring}\label{sec:ex3}

\begin{figure}
    \centering
    \includegraphics[width=0.8\textwidth]{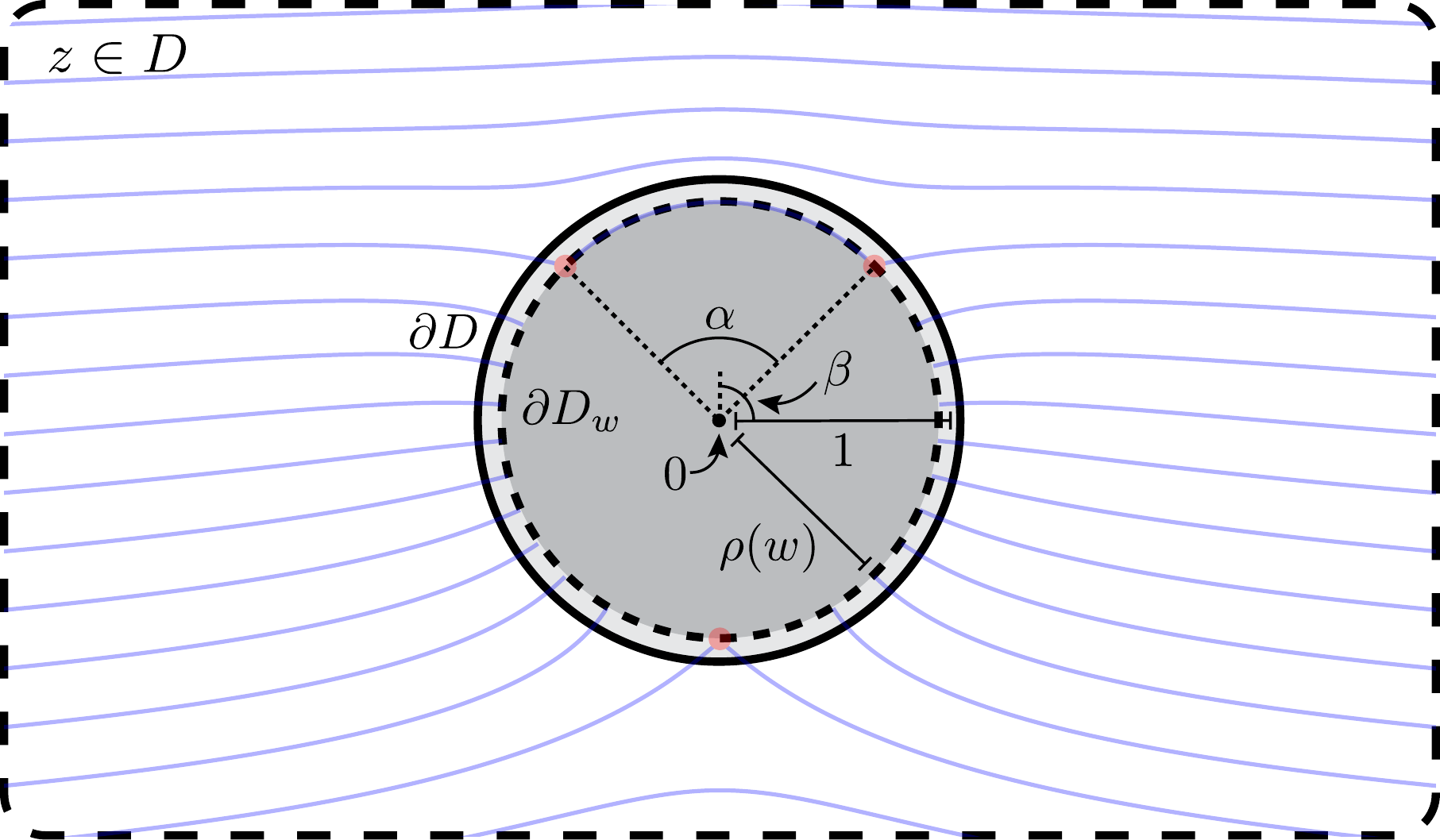}
    \caption{\emph{Example 3.} Two-dimensional liquid crystal outside a unit cylinder ($\partial D$, solid black curve)  subject to weak (finite) tangential anchoring on an arc of angle $\alpha$ and weak normal anchoring on the remaining cylinder. The effective domain boundary is a shrunken cylinder of radius $\rho(w)$ ($\partial D_w$, black dotted curve). The director field inside the effective domain, \eqref{eq:ex3_directorangle}, is shown as faded blue curves for $w=10$, $\alpha=\pi/4$, and an orientation   which minimizes the free energy, $\beta=\pi/2$.} 
    \label{fig:ex3_sketch}
\end{figure}

For a final example, let us reconsider the oriented liquid crystal outside a  cylinder with dimensionless unit radius (as  was considered in \S\ref{sec:ex1}). Unlike the previous two examples, here we assume the liquid crystal is subject to a weak hybrid anchoring condition  (finite $w$), which is partially tangential (homogeneous) and partially perpendicular (homeotropic). That is, we assume  $\Omega(z)$ satisfies \eqref{eq:moderateanchoring} on $|z|=1$ with the anchoring angle
\begin{equation}\label{eq:ex3_hybrid}
    \phi(s) = \arg z'(s) +\begin{dcases}
    0 & \text{if $\left|\arg\left[z(s) \e^{-\im\beta}\right]\right|<\alpha/2$,}\\
     \pi/2 & \text{if $\left|\arg\left[z(s) \e^{-\im\beta}\right]\right|>\alpha/2$,}
    \end{dcases}
    \mod\pi,
\end{equation}
 where $ -\pi<\arg z\leq\pi$ is the principal-value argument and  $\alpha\in[0,2\pi]$ and $\beta\in (-\pi, \pi]$ are given angles. This boundary condition corresponds to imposing tangential anchoring on an arc of angle $\alpha$ and normal anchoring on the remaining circle. The angle $\beta$ corresponds to the orientation of the tangential section of the cylinder. Similar theoretical configurations have focused on Janus particles in three-dimensions \citep{arbzz09,skrd20} and in a two-dimensional film \citep{ls22}; related work on squirming particles may be found by  \cite{lws17,dm18,cgba22,bcpy22}. This configuration is plotted in Fig.~\ref{fig:ex3_sketch}.

Assuming there are no defects within the liquid crystal and it is oriented with the $x$-axis at infinity (\ie~$\Omega(z)\to 0$ as $|z|\to \infty$), $\Omega(z)$ must be single-valued with the period in \eqref{eq:OmegaPeriod} vanishing (\ie~$\Upsilon=M=0$). Furthermore, since both tangential and normal anchoring conditions are conserved under the effective boundary technique (see \S\ref{sec:potential}), imposing finite hybrid anchoring, \ie~\eqref{eq:moderateanchoring} with \eqref{eq:ex3_hybrid}, on $|z|=1$ is equivalent to imposing strong hybrid anchoring, \ie~\eqref{eq:modstronganchoring} with \eqref{eq:ex3_hybrid}, on the effective cylinder $|z|=\rho(w)$, where $\rho(w) \coloneqq (\sqrt{1+4/w^2}-2/w)^{1/2}$ is the effective radius derived in \S\ref{sec:ex1}. Analytically continuing $\Omega(z)$ inside the annulus $\rho(w)<|z|<1$ yields the problem: find a $\Omega(z)$ such that
\begin{subequations}\label{eq:ex3_problem}
\begin{align}
\Omega(z) \text{ holomorphic} &\qquad \text{ in $|z|>\rho(w)$,}\\
 \Im \left[z_s\exp{\Omega(z)}\right]=0 &\qquad \text{ on $|z|=\rho(w)$ with $\left|\arg[z \e^{-\im\beta}]\right|<\alpha/2$,}\\
 \Re \left[z_s\exp{\Omega(z)}\right]=0 &\qquad \text{ on $|z|=\rho(w)$ with $\left|\arg[z \e^{-\im\beta}]\right|>\alpha/2$,}\\
 \Omega(z)\sim  1 &\qquad \text{ as $|z|\to\infty$}.
\end{align}
\end{subequations}

Since $z_s=\im z$ for a circular boundary,  it is convenient to introduce the complex function $F(z) \coloneqq  \im z \exp\Omega(z)$. It follows from \eqref{eq:ex3_problem} that $F(z)$ must be holomorphic inside the extended domain, has a pole at infinity, and has either the real or imaginary part vanishing on  $|z|=\rho(w)$. This problem is conformally invariant, thus we are  free to use  a conformal map to derive a solution.

Consider the  conformal map from the $z$-plane to the $\zeta$-plane
\begin{equation}\label{eq:ex3_conformalmap}
\zeta(z)=\e^{\im\alpha/4} \sqrt{\frac{z-\rho(w) \e^{\im(\beta-\alpha/2)}}{z-\rho(w) \e^{\im(\beta+\alpha/2)}}}.
\end{equation}
This  map is a composition of two conformal maps: a M\"obius transformation, which maps the punctured $z$-plane to the upper-half--plane, and the square-root map, which maps the upper-half--plane to the first quadrant, \ie~the $\zeta$-plane. 

In the $\zeta$-plane, the complex function $G(\zeta)\coloneqq F(z(\zeta))$ satisfies 
\begin{subequations}\label{eq:ex3_zetaproblem}
\begin{align}
G(\zeta) \text{ holomorphic} &\qquad \text{ in $\Re\zeta>0$ \&  $\Im\zeta>0$,}\\
\Im G(\zeta)=0 &\qquad \text{ on $\Im\zeta=0$,}\\
\Re G(\zeta)=0 &\qquad \text{ on $\Re\zeta=0$,}\\
G(\zeta)\sim  \frac{\rho(w)\sin(\alpha/2) \e^{\im\beta}}{1-\e^{-\im\alpha/4} \zeta}  &\qquad \text{ as $\zeta\to  \e^{\im\alpha/4}$}.
\end{align}
\end{subequations}
Here, the pole at $z=\infty$ has been mapped to an interior point  at $\zeta=\e^{\im\alpha/4}$.  The boundary conditions \myeqref{eq:ex3_zetaproblem}{b,c} can be satisfied by introducing three additional poles at the images of  $\zeta=\e^{\im\alpha/4}$ across the real and imaginary axes. This  is commonly referred to as the method of images \citep{af03} and  yields  the solution to \eqref{eq:ex3_zetaproblem},
\begin{equation}\label{eq:ex3_G}
G(\zeta) =  \frac{ \rho\sin(\alpha/2)\e^{\im\beta}}{1-\e^{-\im\alpha/4} \zeta}+\frac{\rho\sin(\alpha/2)\e^{-\im\beta}}{1+\e^{\im\alpha/4} \zeta}-\frac{\rho\sin(\alpha/2)\e^{-\im\beta}}{1-\e^{\im\alpha/4} \zeta}-\frac{\rho\sin(\alpha/2)\e^{\im\beta}}{1+\e^{-\im\alpha/4} \zeta}.
\end{equation}

Changing variables with \eqref{eq:ex3_conformalmap}  results in an expression for the  director angle outside the  cylinder,
\begin{equation}\label{eq:ex3_directorangle}
\begin{split}
\theta(z) &\equiv -\arg\left[-\im F(z)/z\right] =2\arg z-\arg\left(z-\rho \e^{-\im\beta}\right)\\
&\qquad-\frac{1}{2}\arg\left(z-\rho \e^{\im(\beta+\alpha/2)}\right)- \frac{1}{2}\arg\left(z-\rho \e^{\im(\beta-\alpha/2)}\right).
\end{split}
\end{equation}
This solution is equivalent to an unrestricted director field with four defects: two $-1/2$ defects corresponding to the tangential--normal anchoring interfaces at $z=\rho \e^{\im(\beta\pm \alpha/2)}$, one $-1$ defect at $z=\rho \e^{-\im\beta}$, and one $+2$ defect at $z=0$. 

When $\alpha=2\pi$, the liquid crystal is subject to finite tangential anchoring on the entire cylinder (as considered in \S\ref{sec:ex1}). The director angle \eqref{eq:ex3_directorangle}  recovers \eqref{eq:ex1_theta} in this case, with the body orientation, $\beta$,  corresponding to the period  $\gamma\coloneqq \arcsin [\Gamma/(4\pi\rho)]$. Alternatively, when $\alpha=0$, the liquid crystal is  subject to finite normal anchoring on the entire cylinder. The resulting director angle is equivalent to taking $z\mapsto \im z$ in \eqref{eq:ex1_theta} --- \ie~the equipotential lines of the complex potential \eqref{eq:ex1_sol} rotated by $\pi/2$. The body orientation, $\beta$, is again analogous to the period $\gamma$ in this case. Finally, when $\alpha=\pi$, the liquid crystal is subject to finite tangential/normal anchoring on exactly one half of the cylinder.  This boundary condition has previously been termed `Janus anchoring' due to the analogy with Janus particles \citep{ls22}. For strong Janus anchoring ($w=\infty$ and $\alpha=\pi$), the  director angle \eqref{eq:ex3_directorangle} is equivalent to a subcase of the solution derived by \cite{ls22}, who used a superposition of separable solutions to construct a series representation. 

For any given value of $\alpha$, there exists a free energy  associated with each  orientation angle $\beta$, \ie~\eqref{eq:complexenergy}. It is, thus, natural to ask what orientation  minimizes this net free energy. (This  is equivalent to asking what period, $\gamma$, minimizes the net free energies in the previous two examples, \S\ref{sec:ex1} and \S\ref{sec:ex2}.) Since the anchoring angle \eqref{eq:ex3_hybrid} is only defined piecewise, the energy  \eqref{eq:complexenergy} cannot be written as a closed integral of an analytic function and, thus, cannot be evaluated using  Cauchy's residue theorem (as was done in \S\ref{sec:ex1}, when $\alpha=2\pi$, for example). We   instead  evaluate the energy  using numerical quadrature for given values of the anchoring strength, $w$; sector angle, $\alpha$; and orientation angle, $\beta$. We are then able to  minimize over the angle $\beta$ numerically. Further details on the evaluation of  the energy integral, \eqref{eq:complexenergy}, along with some partial analytical progress can be found in App.~\ref{app:energy_ex3}.

\begin{figure}
    \centering
    \includegraphics[width=.9\textwidth]{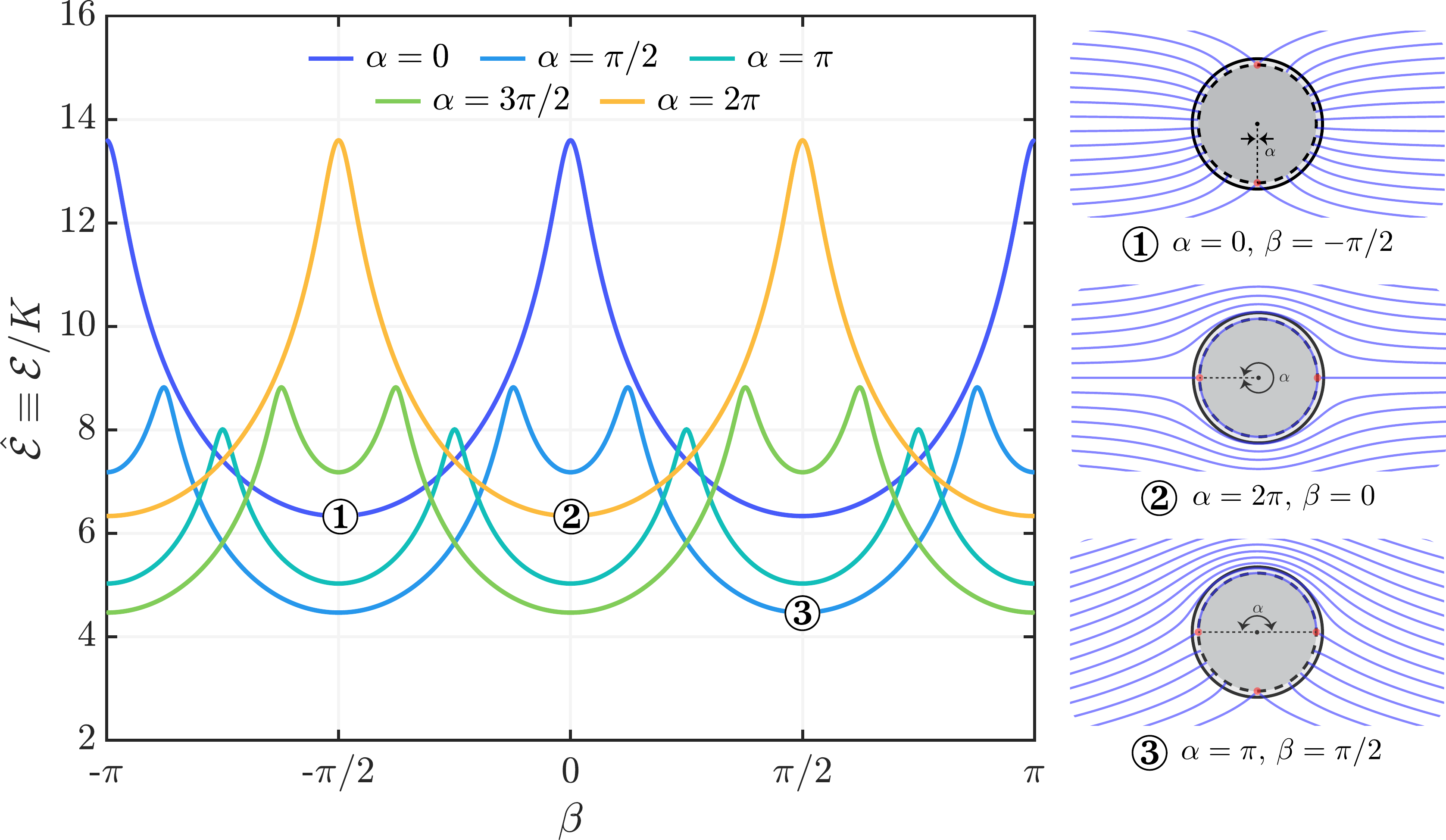}
    \caption{\emph{Example 3.} Plot of the net free energy, \eqref{eq:complexenergy}, as a function of the  orientation angle, $\beta$, for anchoring strength $w=10$ and  various sector angles, $\alpha$, delineated by colour in the figure legend. It is evident that  $\beta=-\pi/2$ and  $\beta=\pi/2$   minimize the net free energy  when $0\leq\alpha< \pi$  (as shown in \ding{172}, for example), whilst $\beta=0$ and $\beta=2\pi$   minimize the net free energy when $\pi<\alpha\leq 2\pi$  (as shown in \ding{173}, for example). In the critical case ($\alpha=\pi$), all four orientation  angles  minimize the net free energy (as shown in \ding{174}, for example). } 
    \label{fig:ex3_Energy}
\end{figure}

If the anchoring  is predominately normal to the cylinder (\ie~$0\leq\alpha<\pi$), we observe that the net free energy is minimized for a vertically-aligned cylinder (\ie~$\beta=\pm\pi/2$).  Alternatively, if the anchoring  is predominately tangential to the cylinder (\ie~$\pi<\alpha\leq 2\pi$), then the net free energy is minimized by a horizontally-aligned cylinder (\ie~$\beta=0$ or $2\pi$). In the critical case (\ie~Janus anchoring, $\alpha=\pi$), both horizontal and vertical alignment minimize the free energy. This observation is supported by Fig.~\ref{fig:ex3_Energy}, which shows the net free energy as a function of the orientation angle, $\beta$, for various values of $\alpha$ and $w=10$. It also complements the analytical expression of the energy \eqref{eq:ex1_AnalyticEnergy}, which holds when $\alpha=2\pi$. This orientation preference suggests that, if the cylindrical body was free to rotate, it would align itself with either the vertical or horizontal axes depending on the value of $\alpha$. Controlling the orientation of a body using surface anchoring has previously been studied in the context of microswimmers by \cite{cpzba20}, and a related method of fluid transport by passing a traveling wave of preferred director angle has elsewhere been proposed by \cite{ksp15}. However, analysing such  orientation control further requires the addition of dynamics within the liquid crystal and, thus, is left for future work.

\section{Conclusions} \label{sec:conc}

We have presented a path towards analytical solutions for the equilibrium nematic director field with an inclusion in two-dimensions. A complex variables approach was found to provide a comfortable means of determining not only the director field but the surface tractions, net force, and net torque on the immersed body. These quantities are not immediately accessible by other means due to topological conservation laws and the nonlinear boundary condition associated with weak (finite) anchoring constraints. One remaining question pertains to the accuracy of the asymptotic solutions derived for large anchoring strengths ($w\gg1$) using the effective boundary technique. Although not included here, preliminary comparisons of the analytical results described here with full numerical simulations suggest that the director field and even local tractions are predicted with high accuracy even for small anchoring strengths.

The analytical solutions provided here may represent a starting point to advance other calculations, for instance when calculating viscous drag on an immersed particle. The force and torque on particles with a non-uniform background director field might also be determined by adjusting the theory presented here. A non-trivial body force and torque, or alternatively particle migration and rotation, are expected in that setting. Equilibrium states also provide a basis for understanding the LC-mediated elastic interactions between two bodies, and the deformation of soft immersed bodies like red blood cells which `share the strain' with the bulk liquid crystal \citep{nesa20}. These extensions will be explored in greater detail in follow-up work.


\backsection[Acknowledgements]{The authors acknowledge helpful conversations with Nicholas L. Abbott and Thomas R. Powers.}

\backsection[Funding]{The research leading to these results has received funding from the NSF grant DMR-2003819.}

\backsection[Declaration of interests]{The authors report no conflict of interest.}


\backsection[Author ORCIDs]{T.~G.~J.~Chandler, https://orcid.org/0000-0001-5351-5399; S.~E.~Spagnolie, https://orcid.org/0000-0002-8484-1573}


\appendix

\section{Body forces and torques}\label{app:variational}
In this appendix, we  derive expressions for the forces and torques associated with the net free energy
\begin{equation}\label{eq:netenergy}
    \mathcal{E} = \int_{D} \F(\nabla\theta)\de A + \int_{\partial D} \F^s(\theta,\phi)\de s,
\end{equation}
where $\F(\nabla\theta)\coloneqq K|\nabla\theta|^2/2$ and $\F^s(\theta,\phi)\coloneqq W\sin(\theta-\phi)^2/2$. We shall do this by applying a virtual work argument, following similar derivations by  \cite{degennes1993,stewart2004,virga2018}. Throughout, we denote the boundary tangent as $\bm{\s}\coloneqq\bm{x}_s$ and the liquid crystal--pointing normal as $\bm{\n}\coloneqq-\bm{x}_s^{\perp}$. Moreover, the anchoring angle, $\phi$, is  assumed to be a function of the local boundary orientation: for example, homogeneous anchoring, \ie~$ (\cos\phi,\sin\phi)\parallel \bm{\s}$, and homeotropic  anchoring, \ie~$(\cos\phi,\sin\phi) \parallel  \bm{\n}$.

\subsection{Variation of the bulk term}

Consider the infinitesimal variation of the position and director angle within the liquid crystal \nolinebreak
\begin{subequations}\label{eq:variations}
\begin{align}
    \bm{x} \quad&\mapsto\quad \bm{x}+\bm{u}(\bm{x}),\\
  \text{and}\quad\theta(\bm{x}) \quad&\mapsto\quad \theta(\bm{x})+\Psi(\bm{x}),
\end{align}
\end{subequations}
respectively. Here, we shall restrict our attention to incompressible variations, \ie~$\nabla\cdot\bm{u}=0$, thus $\de A$ is conserved. Under these variations, the partial derivatives are mapped by 
\begin{subequations}\label{eq:partial_variations}
\begin{align}
     \partial_i\coloneqq \partial_{x_i} \quad&\mapsto\quad  (\delta_{i j}- u_{j,i})\partial_j= \partial_i-u_{j,i}\partial_j,\\
     \partial_i\theta \quad&\mapsto\quad  \theta_{,i}+\Psi_{,i} -\theta_{,j} u_{j,i},
\end{align}
\end{subequations}
where repeated indices imply summation and commas within the index notation denote partial derivatives.

Consider an arbitrary area contained within the liquid crystal bulk,  $\mathcal{D}\subset D$. By the principle of virtual work, variations of the free energy over $\mathcal{D}$ must balance  body and surface forces, $\bm{F}$ and $\bm{f}$, and  generalized body and surface forces, $G$ and $g$, that is
\begin{equation}\label{eq:virtualwork}
   \delta \int_{\mathcal{D}} \F(\nabla\theta)\de A-\int_\mathcal{D} p(\bm{x})\nabla\cdot\bm{u}\de A= 
\int_{\mathcal{D}}\bm{F}\cdot \bm{u} + G\Psi\de V + \int_{\partial \mathcal{D}} \bm{f}\cdot\bm{u} + g\Psi \de S,
\end{equation}
where $\delta$ denotes the variation under \eqref{eq:variations} and   the pressure $p(\bm{x})$ is a Lagrange multiplier, which imposes incompressiblity (\ie~$\nabla\cdot\bm{u}=0$). 

Using \eqref{eq:partial_variations} and the divergence theorem, we find that 
\begin{equation}
\begin{split}
&\delta\int_{\mathcal{D}} \F(\nabla\theta) \de A= \int_{\mathcal{D}} \F_{\theta_j}\cdot(\Psi_{,j} -\theta_{,i} u_{i,j}) \de A \\
&\qquad=  -\int_{\mathcal{D}} \left(\F_{\theta_j}\right)_{,j}\Psi - \left(\F_{\theta_j} \theta_{,i}\right)_{,j} u_{i}\de A - \int_{\partial \mathcal{D}} \F_{\theta_j}\n_j\Psi - W_{\theta_j}\n_j \theta_{,i}u_{i}\de s.
\end{split}
\end{equation}
 By defining $\sigma^d_{ij} \coloneqq - \F_{\theta_j}\theta_{,i}$ and $M_j\coloneqq \F_{\theta_j}$, this can be written as
\begin{equation}
\delta\int_{\mathcal{D}} \F(\nabla\theta) \de A=  -\int_{\mathcal{D}} M_{j,j}\Psi +\sigma^d_{ij,j} u_{i}\de A - \int_{\partial \mathcal{D}} M_j\n_j\Psi +\sigma^d_{ij}\n_ju_{i}\de s.
\end{equation}
The Lagrange multiplier term can be written as
\begin{equation}
-\int_{\mathcal{D}}p(\bm{x}) u_{i,i}\de A \equiv \int_{\mathcal{D}}p_{,i} u_{i}\de A +\int_{\partial \mathcal{D}}p \n_iu_{i}\de s,
\end{equation}
by the divergence theorem. Overall, we find that
\begin{equation}\label{eq:bulkcontribution}
\begin{split}
   &\delta \int_{\mathcal{D}} \F(\nabla\theta)\de A-\int_\mathcal{D} p(\bm{x})\nabla\cdot\bm{u}\de A\\
   &\qquad=  -\int_{\mathcal{D}} M_{j,j}\Psi +\sigma_{ij,j}^Eu_{i}\de A - \int_{\partial \mathcal{D}} M_j\n_j\Psi +\sigma_{ij}^E\n_j u_{i}\de s,
   \end{split}
\end{equation}
where we define $\sigma_{ij}^E \coloneqq -p\delta_{ij}+\sigma^d_{ij}$.

We are now ready to balance the energy variation with the virtual work, as in \eqref{eq:virtualwork}. From the translation variation, \ie~$\bm{u}$, we obtain the force equilibrium equations \nolinebreak
\begin{subequations}\label{eq:force_equilbrium}
\begin{equation}
F_i + \sigma^E_{ij,j}=0 \quad \text{and} \quad t_i +\sigma^E_{ij}\n_j = 0,\subtag{a,b}
\end{equation}
\end{subequations}
in $\mathcal{D}$ and on $\partial \mathcal{D}$, respectively. We can, thus, interpret $\sigma_{ij}^E=-p\delta_{ij} - \F_{\theta_j}\theta_{,i}$ as a stress tensor (\ie~the Ericksen stress tensor). From the rotational variation, $\ie~\Psi$, we obtain the moment equilibrium equations
\begin{subequations}\label{eq:moment_equilbrium}
\begin{equation}
G+ M_{j,j}=0  \quad \text{and} \quad g + M_j\n_j = 0,\subtag{a,b}
\end{equation}
\end{subequations}
in $\mathcal{D}$ and on $\partial \mathcal{D}$, respectively. We can, thus, interpret $M_j\coloneqq \F_{\theta_j}$ as a generalized stress vector.

Assuming there are no bulk forces (\ie~$F_i=G=0$), the equilibrium equations, \myeqref{eq:force_equilbrium}{a} and \myeqref{eq:moment_equilbrium}{a}, take the form
\begin{subequations}\label{eq:equilbriumeq_noforce}
\begin{equation}
\sigma^E_{ij,j}\equiv-(p\delta_{ij}+ \F_{\theta_j}\theta_{,i})_{,j}=0 \quad \text{and} \quad M_{j,j}\equiv\left(\F_{\theta_j}\right)_{,j}=0,\subtag{a,b}
\end{equation}
\end{subequations}
in $\mathcal{D}$. By the chain rule, $(\F_{\theta_j}\theta_{,i} )_{,j}\equiv (\F_{\theta_j})_{,j}\theta_{,i}+\F_{\theta_j}\theta_{,ji}\equiv (\F_{\theta_j})_{,j}\theta_{,i}+\F_{,i}$, and thus \eqref{eq:equilbriumeq_noforce} is directly equivalent to
\begin{subequations}\label{eq:equilbriumeq_noforce2} 
\begin{equation}
(p+ \F)_{,i}=0 \quad \text{and} \quad  \left(\F_{\theta_j}\right)_{,j}=0,\subtag{a,b}
\end{equation}
\end{subequations}
in $\mathcal{D}$. Integrating \myeqref{eq:equilbriumeq_noforce2}{a}, yields an expression for the pressure $p=p_0-\F$ for an arbitrary constant $p_0$, leaving \myeqref{eq:equilbriumeq_noforce2}{b} as the final equilibrium equation. (Note  that a similar result holds if the bulk forces are conservative.) To determine the physical boundary conditions, we now consider the case when $\mathcal{D}=D$.

\subsection{Variation of the surface term}  
When applying the principle of virtual work on the domain $D$, one will  obtain a  contribution from the surface energy term in \eqref{eq:netenergy},  on top of the already derived  bulk contribution in \eqref{eq:bulkcontribution}. In this case, the surface force, $\bm{f}$, and generalized surface force, $g$, must satisfy
\begin{equation}\label{eq:surfacetermvariation1}
\delta\int_{\partial D} \F^s(\theta,\phi)\de s- \int_{\partial D} M_j\n_j\Psi +\sigma_{ij}^E\n_j u_{i}\de s = \int_{\partial D} \bm{f}\cdot\bm{u} + g\Psi \de S,
\end{equation}
where the first and second integrals are the contribution from the surface and bulk energy, respectively.

On the domain boundary, $\partial D$, we  consider a translation and rotation variation of the same form as \eqref{eq:variations}. Under these variations, we find that
\begin{subequations}\label{eq:surfacevariations}
\begin{align}
    \de s \quad&\mapsto\quad   |\bm{x}_s+\bm{u}_s|\de s \sim(1 + \bm{\s}\cdot\bm{u}_{s})\de s,\\
    \bm{\s} \coloneqq \bm{x}_s \quad&\mapsto\quad  (\bm{x}+\bm{u})_s/(1 + \bm{u}_s\cdot\bm{\s}) \sim \bm{\s}  + (\bm{\n}\cdot\bm{u}_{s})\bm{\n},\\
        \bm{\n} \coloneqq -\bm{x}_s^{\perp}\quad&\mapsto\quad  -(\bm{x}^\perp+\bm{u}^\perp)_s/(1 + \bm{u}_s\cdot\bm{\s}) \sim \bm{\n} -(\bm{\n}\cdot\bm{u}_{s})\bm{\s},
        \\
    \phi \quad&\mapsto\quad   \phi -\bm{\n}\cdot\bm{u}_{s},
\end{align}
\end{subequations}
where subscript $s$ denotes an arclength derivative. Note that the variation in $\phi$, \ie~\myeqref{eq:surfacevariations}{d}, comes from the assumption that $\phi$ is a function of the local boundary angle. 

Using \eqref{eq:surfacevariations}, we find that
\begin{equation}\label{eq:surfacetermvariation2}
\begin{split}
 &\delta\int_{\partial \tilde{D}} \F^s(\tilde{\theta},\tilde{\phi})\de \tilde{s} = \int_{\partial D} \F^s_{\theta}\Psi -\F^s_\phi\cdot(\bm{\n}\cdot\bm{u}_{s})+\F^s\cdot( \bm{\s}\cdot\bm{u}_{s})\de s\\
 &\quad=\int_{\partial D} \F^s_{\theta}\Psi +\left(\F^s_\phi\bm{\n}-\F^s \bm{\s} \right)_{s} \cdot\bm{u}\de s-\left[( \F^s_\phi\bm{\n}-\F^s \bm{\s} ) \cdot\bm{u}\right]_{\partial D}.
 \end{split}
\end{equation}
Here, the final term comes from integrating by parts and corresponds to a net jump around the boundary $\partial D$. For a closed boundary, this term will vanish; thus,  we shall  disregard it.

Combining \eqref{eq:surfacetermvariation1} and \eqref{eq:surfacetermvariation2} yields expressions for the surface force, $\bm{f}$, and generalized surface force, $g$. From the translational variation, $\ie~\bm{u}$, we obtain the surface force
\begin{equation}\label{eq:app_feq}
    -\bm{f} = (\F-p_0)\bm{\n} -\F_{\theta_j}\n_j\nabla\theta+(\F^s \bm{\s} -\F^s_\phi\bm{\n})_{s},
\end{equation}
where we have used the fact that $\sigma^E_{ij}\n_j\equiv (\F-p_0)\n_i -\F_{\theta_j}\n_j\theta_{,i}$. From the rotational variation, $\ie~\Psi$, we obtain the generalized surface force
\begin{equation}\label{eq:app_geq}
    -g = -\F^s_{\theta} + \F_{\theta_j}\n_j,
\end{equation}
where we have used  the fact that $M_j = \F_{\theta_j}$.

It is natural to assume zero surface moment  (\ie~$g=0$) at the boundary; thus, from \eqref{eq:app_geq}, we obtain the boundary condition
\begin{equation}\label{eq:equilbriumbc}
 \F_{\theta_j}\n_j  -\F^s_{\theta} = 0,
\end{equation}
on $\partial D$.

Overall,  \eqref{eq:equilbriumeq_noforce2} and \eqref{eq:equilbriumbc} yield the equilibrium equation
\begin{equation}\label{eq:app_equileq}
\left(\F_{\theta_j}\right)_{,j}=0  \quad \text{subject to}\quad 
 \F_{\theta_j}\n_j  =\F^s_{\theta},
\end{equation}
with  the surface traction on $\partial D$  given by \eqref{eq:app_feq}.

\subsection{Dirichlet with Rapini--Papoular  energy} 

For the Dirichlet bulk energy, $\F\coloneqq K|\nabla\theta|^2/2$, and  Rapini--Papoular surface energy, $\F^s\coloneqq W\sin(\theta-\phi)^2/2$, considered in this paper, the equilibrium equations \eqref{eq:app_equileq} take the form
\begin{subequations}\label{eq:finalequilbrium}
\begin{equation}
\nabla^2\theta =0 \quad \text{subject to} \quad K\frac{\partial \theta}{\partial\nu}= \frac{W}{2}\sin\left[2(\theta-\phi)\right],\subtag{a,b}
\end{equation}
\end{subequations}
whilst the surface traction, \eqref{eq:app_feq}, is
\begin{equation}\label{eq:finaltraction}
      -\bm{f}=\frac{K}{2}|\nabla\theta|^2\bm{\n}-K\frac{\partial \theta}{\partial\nu} \nabla\theta+ \frac{W}{2}\left\{\sin\left[2(\theta-\phi)\right]\bm{\n} +\sin(\theta-\phi)^2 \bm{\s} \right\}_{s},
\end{equation}
up to an additive constant pressure, $-p_0\bm\n$. For homogeneous or homeotropic anchoring, $\bm{\n}_s= \phi_s\bm{\s}$ and $\bm{\s}_s=-\phi_s\bm{\n}$; consequently, one can show that the traction, \eqref{eq:finaltraction}, acts purely in the normal direction, \ie
\begin{equation}
      -\bm{f}=\left[\frac{K}{2}\left(\theta_s^2-\theta_{\nu}^2\right)+K\theta_{\nu s}-\frac{W}{2}\sin(\theta-\phi)^2 \phi_s\right]\bm{\n} ,
\end{equation}
where the boundary condition \myeqref{eq:finalequilbrium}{b} has been imposed.

\section{Computing free energies}

In this appendix, we provide further details on the evaluation of the complex energy integral, \eqref{eq:complexenergy}, for the three examples presented in the main text. We begin by noting that $\Omega(z)$ is single-valued in all three examples, thus the energy \eqref{eq:complexenergy} can be written as the complex contour integral
\begin{equation}\label{eq:app_energy}
\hatE = -\frac{1}{4}\Im\left[\oint_{\partial D} \overline{\Omega( z)} \Omega'(z)\de z\right]+ \frac{w}{4} \Re\left[P-\oint_{\partial{D}}\e^{\Omega(z)-\overline{\Omega( z)}}\frac{\e^{2\im\phi}}{z'(s)} \de z\right],
\end{equation}
where $P$ is the perimeter of $\partial D$.

\subsection{Example 1}\label{app:energy_ex1}

In the first example, $\partial D$ denotes the unit circle, $|z|=1$;  the anchoring angle, $\phi$, was defined such that $\e^{\im\phi} = z'(s) = \im z$; and  the complex potential, $f(z)$, was defined such that $\Omega(z)=\log f'(z)$. Inserting these expressions, along with the unit circle's Schwarz function, $\bar{z}=1/z$, and $P=2\pi$ into \eqref{eq:app_energy} yields  the contour integral given in the main text, \ie~\eqref{eq:ex1_energycontour}.

The complex potential was found to be \eqref{eq:ex1_sol}, thus we have that
\begin{subequations}\label{eq:app_ex1_df}
\begin{align}
    f'(z) &= (z- \rho \e^{\im\gamma})(z+\rho \e^{-\im\gamma})/z^2,\\
\text{and}\quad\overline{f}'(1/z) &=-\rho^2 (z-\e^{-\im\gamma}/\rho) (z+\e^{\im\gamma}/\rho),
\end{align}
\end{subequations}
where $\sin\gamma\coloneqq \Gamma/(4\pi\rho)$. It is apparent from \eqref{eq:app_ex1_df} that $f'(z)$ has two zeros at $z=\pm \rho \e^{\pm\im\gamma}$  and one pole at $z=0$ (\ie~the three topological defects), whilst $\bar{f}'(1/z)$ only has two zeros at $z=\pm \e^{\pm\im\gamma}/\rho$. Consequently, the integrand  in  \eqref{eq:ex1_energycontour} is holomorphic in $|z|\leq 1$ except for three simple poles at $z= \rho \e^{\im\gamma}$, $z=- \rho \e^{-\im\gamma}$, and $z=0$. The analytical expression \eqref{eq:ex1_AnalyticEnergy} follows from applying  Cauchy's residue theorem.

\subsection{Example 2}\label{app:energy_ex2}

In the second example, $\partial D$ denotes the equilateral triangle with corners at the roots of $z^3=\e^{3\im\chi}$, whilst the anchoring angle, $\phi$, and complex potential, $f(z)$, are again  defined such that $\e^{\im\phi} = z'(s)$ and $\Omega(z)=\log f'(z)$, respectively. Inserting these expressions and $P=3\sqrt{3}/2$ into \eqref{eq:app_energy} yields
\begin{equation}\label{eq:app_ex2_energy}
\hatE =\frac{3\sqrt{3}}{8}w -\frac{1}{4}\oint_{\partial D}\Im\left[ \log\overline{f'(z)} \frac{f''(z)}{f'(z)}z'(s)\right]+ w \Re\left[\frac{f'(z)}{\overline{f'(z)}}z'(s)^2\right]\de s,
\end{equation}
where 
\begin{equation}\label{eq:app_ex2_f}
f'(z) = \frac{\left(\zeta(z)-\e^{\im\gamma}\right)\left(\zeta(z)+\e^{-\im\gamma}\right)}{(\zeta(z)^3-\e^{\im\chi})^{2/3}},
\end{equation}
is the potential derived in the main text, $\sin\gamma\coloneqq h(1)\Gamma/[4\pi(1-2/w)]$, and $\zeta(z)$ is given by the inverse of \eqref{eq:ex2_zzeta}.

To evaluate the integral in \eqref{eq:app_ex2_energy}, we  split the integration contour into three line segments corresponding to the three sides of the triangle, that is $\partial D = L_1\cup L_2\cup L_3$ where $L_1 \coloneqq [\e^{\im\chi},\e^{\im(\chi+2\pi/3)}]$, $L_2 \coloneqq [\e^{\im(\chi+2\pi/3)},\e^{\im(\chi-2\pi/3)}]$, and  $L_3 \coloneqq [\e^{\im(\chi-2i\pi/3)},\e^{\im\chi}]$. These segments can be individually parameterized as
\begin{equation}\label{eq:app_ex2_par}
z(s) = \e^{\im\chi}\begin{cases}
1-\e^{-\im\pi /6}s &\quad \text{for $L_1$,}\\
\e^{2i\pi/3} -\im s&\quad \text{for $L_2$,}\\
\e^{-2i\pi /3}+\e^{\im\pi /6}s&\quad \text{for $L_3$,}\\
\end{cases}
\end{equation}
where $s\in[0, \sqrt{3}]$ is  the corresponding arclength. The arclength is then discretized and the corresponding integrals are evaluated numerically for given $\chi$, $\gamma$, and $w$ --- we use numerical quadrature using \textsc{Matlab}. 

Note that the Schwarz--Christoffel mapping in  \eqref{eq:ex2_zzeta} must be inverted numerically in order to evaluate \eqref{eq:app_ex2_f}. This is known to be numerically challenging \citep{dt02}. To alleviate this issue, we use the Schwarz--Christoffel toolbox by \cite{driscoll1996}.

\subsection{Example 3}\label{app:energy_ex3}
In the third example, $\partial D$ denotes the unit circle, $|z|=1$, and the anchoring angle, $\phi$, is given by \eqref{eq:ex3_hybrid} with $z'(s)=\im z$. Inserting these expressions along with the unit circle's Schwarz function, $\bar{z}=1/z$, and $P=2\pi$ into \eqref{eq:app_energy} yields  the contour integral
\begin{equation}\label{eq:app_ex3_energy}
\hatE = \frac{\pi}{2}w +\frac{1}{4}\Im\left[-\oint_{|z|=1} \overline{\Omega}( 1/z) \Omega'(z)\de z+w\left(\int_{\partial D_1}-\int_{\partial D_2}\right)\e^{\Omega(z)-\overline{\Omega}(1/ z)}z \de z\right],
\end{equation}
where $\partial D_1$ and $\partial D_2$ denote the segments of $\partial D$ which are subject to tangential anchoring, $\left|\arg\left[z \e^{-\im \beta}\right]\right|<\alpha/2 $, and normal anchoring, $\left|\arg\left[z \e^{-\im \beta}\right]\right|>\alpha/2$,  respectively. 

The complex director angle, $\Omega(z)$, was determined in the main text to be
\begin{equation}\label{eq:app_ex3_omega}
   \e^{\Omega(z)}= z^{-2}\left(z-\rho \e^{-\im\beta}\right) \sqrt{\left(z -\rho \e^{\im(\beta+\alpha/2)}\right)\left(z-\rho \e^{\im(\beta-\alpha/2)}\right)},
\end{equation}
where the complex square root is defined  such that 
 it is of principal value with its  branch cut spanning the arc  $z\in[\rho \e^{\im(\beta-\alpha/2)}, \rho \e^{\im(\beta+\alpha/2)}]$.

We shall begin by considering the first integral in \eqref{eq:app_ex3_energy}. Since $0\leq\rho<1$, it follows immediately from \eqref{eq:app_ex3_omega} that $\bar{\Omega}(1/z)$ is holomorphic inside $|z|=1$. We, hence, find that the integrand,  $\bar{\Omega}(1/z)\Omega'(z)$, is holomorphic in $|z|\leq 1$ except for three simple poles. These poles correspond to the three  defects on the effective cylinder:  $z=\rho \e^{-\im\beta}$, $z= \rho \e^{\im(\beta+\alpha/2)}$, and $z=\rho \e^{\im(\beta-\alpha/2)}$. (Note that the fourth defect, $z=0$, is a removable singularity and so can be ignored.) It, thus, follows from Cauchy's residue theorem that
\begin{equation}
\begin{split}
  \Im\left[  \oint_{|z|=1} \overline{\Omega}( 1/z) \Omega'(z)\de z \right]= 3\pi\log\left|1-\rho^2\right|+\pi\log\left|1-\rho^2 \e^{\im \alpha}\right|&\\
+
2\pi\log\left|1-\rho^2  \e^{\im(2\beta+\alpha/2)}\right|+2\pi\log\left|1-\rho^2  \e^{\im(2\beta-\alpha/2)}\right|&.
\end{split}
\end{equation}

The latter two integrals in \eqref{eq:app_ex3_energy} can not be evaluated using Cauchy's residue theorem since the contours reside on different Riemann sheets of the integrand. Instead we evaluate the integrals  numerically by discretizing the unit circle and applying quadrature in \textsc{Matlab}. Note that extra care must be taken to ensure the principal value square root is used here.

\bibliographystyle{jfm}
\bibliography{jfm}






\end{document}